\documentclass[onecolumn,floats]{revtex4}

\usepackage{epsfig}
\usepackage{amssymb,latexsym,amsmath}
\newcommand{\eq}[1]{\begin{align} #1 \end{align}}

\begin{document}

\title{Fluctuations in the Canonical Ensemble}

\author{ V.V. Begun$^{a}$,  M.I. Gorenstein$^{a,b}$,
O.S. Zozulya$^{a,c}$}

\affiliation{
 $^{a}$ Bogolyubov Institute for Theoretical Physics, Kiev, Ukraine\\
 $^{b}$ Frankfurt Institute for Advanced Studies, Frankfurt, Germany\\
 $^{c}$ Utrecht University, Utrecht, The Netherlands}

\begin{abstract}
The particle number and energy fluctuations
in the system of charged particles are studied in the canonical
ensemble for non-zero net values of the conserved charge.
In the thermodynamic limit the fluctuations in
the canonical ensemble are  different from the fluctuations in the grand
canonical one. The system with several species of particles is considered.
We calculate the quantum statistics effects
which can be taken into account for the canonical ensemble fluctuations
in the infinite volume limit.
The fluctuations of the particle numbers in the pion-nucleon gas are
considered in the canonical ensemble as
an example of the system with two conserved
charges -- baryonic number and electric charge.
\end{abstract}

\maketitle

\section { Introduction}
The statistical model approach turns out to be rather successful
in describing the data on the particle production in relativistic
nuclear collisions (see e.g. Ref.\,\cite{stat-model} and recent
review \cite{PBM}). This stimulates further investigation of the
properties of the statistical models. In particular, an
applicability of various statistical ensembles is an interesting
issue. The canonical ensemble (CE) \cite{ce} or even the
microcanonical ensemble (MCE) \cite{mce} have been used in
order to describe the $pp$, $p\bar{p}$ and $e^+e^-$ 
collisions when a small number of secondary
particles are produced.  At these conditions the
 statistical systems are far away
from the thermodynamic limit, so that the statistical ensembles
are not equivalent, and the exact charge or both energy and charge
conservation laws have to be taken into account.
 The grand canonical ensemble (GCE)
formulation is valid when the system volume $V$ tends to infinity.
All statistical ensemble become then thermodynamically equivalent.
 
 The analysis of the fluctuations is a useful tool to study the properties
  of the system created during  high energy
  particle and nuclear collisions (see e.g. Refs.\,\cite{fluc,fluc1,steph}).
An essential part of the total fluctuations measured on the
event-by-event basis is expected to be the thermal ones.
The particle number fluctuations have been recently studied
 in the CE \cite{ce-fluc} and MCE \cite{mce-fluc} and compared with those
 in the GCE.
  It has been shown that these fluctuations 
  are different in various statistical ensembles 
  in the particular case of the
  relativistic ideal gas with a total net charge equal zero in
  the Boltzmann  statistics  approximation.
The fluctuations of negatively
 and positively  charged particles are 
 suppressed in the CE \cite{ce-fluc} in
comparison to the fluctuations in the GCE.
This suppression remains valid in the thermodynamic limit
too, so that the well-known equivalence of all statistical
ensembles refers to the average quantities, but does not apply to
the fluctuations.

  In Ref.\,\cite{ce-fluc} we have studied
  the CE for one particle specie and zero
 net value of the conserved charge. In the present paper we extend
  our consideration. In the high energy
   proton-proton and
   nucleus-nucleus collisions the created system has some positive
   values of the baryonic number and electric charge. Besides, a
   lot  of different
   species of hadrons are created.  We
  study the CE particle number fluctuations (Secs.\,II and III)
  and energy fluctuations (Sec.\,IV)
  in the systems with non-zero net charge and several species of particles.
   As the electric charge of hadrons can be both $\pm1$ and $\pm2$,
   we consider the CE
  system of 
  single and double  charged particles in Sec\,V.
  The effects of Bose and Fermi statistics
  are studied in the thermodynamic
  limit in Sec.\,VI.
  We also calculate in Sec.\,VII
  the CE particle number fluctuations for the ideal
  pion-nucleon gas  which is an example of the system with two conserved
charges -- baryonic number and electric charge.
We summarize our consideration and formulate the conclusions
in Sec.\,VIII.

\section{The GCE and CE partition functions and mean particle
multiplicities}

 Let us start with the multi-species system of $+1$
and $-1$ charged particles. In applications of the statistical
approach to hadron production in high energy collisions the
conserved charge under consideration can be the electric charge
and baryonic number, or strangeness and charm, which are also
conserved in the strong interactions.
In the case of the Boltzmann ideal gas (i.e. the interactions and quantum
statistics effects are neglected) the partition function in the
GCE reads:

\begin{align}\label{Zgce}
&Z_{g.c.e.}(V,T,\mu)
\;=\;\sum_{N_{1+},~N_{1-}=0}^{\infty}...\sum_{N_{j+},~N_{j-}=0}^{\infty}...
 ~~\frac{\left(\lambda_{1+}z_1\right)^{N_{1+}}}{N_{1+}!}~
 \frac{\left(\lambda_{1-}z_1\right)^{N_{1-}}}{N_{1-}!}~...
~\frac{\left(\lambda_{j+}z_j\right)^{N_{j+}}}{N_{j+}!}~
 \frac{\left(\lambda_{j-}z_j\right)^{N_{j-}}}{N_{j-}!}~...~\nonumber \\
 &=~\prod_{j}~\sum_{N_{j+},~N_{j-}=0}^{\infty}
 \frac{\left(\lambda_{j+}z_j\right)^{N_{j+}}}{N_{j+}!}~
 \frac{\left(\lambda_{j-}z_j\right)^{N_{j-}}}{N_{j-}!} \
=\; \prod_j~\exp\left(\lambda_{j+}z_j~+~\lambda_{j-}z_j\right)~
=~\exp\left[~2z~\cosh\left(\frac{\mu}{T}\right)\right]~,
\end{align}
where $j$ numerates the spesies, $\lambda_{j\pm}=\exp(\pm\mu/T)$, $z_j$ 
is a single particle
partition function
\begin{align}\label{z}
z_j\;=\; \frac{g_jV}{2\pi^2}
       \int_{0}^{\infty}k^{2} dk\;
       \exp\left[-~\frac{(k^{2}+m_j^{2})^{1/2}}{T}\right]
        \;=\; \frac{g_jV}{2\pi^2} \;\;
       T\,m_j^2\,K_2\left(\frac{m_j}{T}\right)~,
\end{align}
and $z\equiv \sum_{j}z_j$.
The $V$, $T$ and $\mu$ are respectively the system volume, temperature and
chemical potential connected with the conserved charge $Q$. The $g_j$ and
$m_j$ are respectively the degeneracy factors and masses for the $j$-th
particle species, and $\;K_2\;$ is the modified Hankel function.
 The CE  partition function is obtained by an
explicit introduction of the charge conservation constrain,
$\sum_{j}\left(N_{j+} - N_{j-}\right) = Q\;$, for each microscopic
state of the system:
\begin{align}\label{Zce}
&Z_{c.e.}(V,T,Q)
~=~\sum_{N_{1+},~N_{1-}=0}^{\infty}...\sum_{N_{j+},~N_{j-}=0}^{\infty}...
 ~~\frac{\left(\lambda_{1+}z_1\right)^{N_{1+}}}{N_{1+}!}~
 \frac{\left(\lambda_{1-}z_1\right)^{N_{1-}}}{N_{1-}!}~...
\frac{\left(\lambda_{j+}z_j\right)^{N_{j+}}}{N_{j+}!}
 \frac{\left(\lambda_{j-}z_j\right)^{N_{j-}}}{N_{j-}!}...\nonumber\\
&\times \delta\left[\left(N_{1+}+...+N_{j+}
+...-N_{1-}-...-N_{j-}-...\right)-Q\right]
~=\;\int_0^{2\pi}\frac{d\phi}{2\pi}~ \prod_{j}
\sum_{N_{j+},N_{j-}=0}^{\infty}\;
 \frac{\left(\lambda_{j+} z_j\right)^{N_{j+}}}{N_{j+}!}
 \frac{\left(\lambda_{j-} z_j\right)^{N_{j-}}}{N_{j-}!}\nonumber \\
&\times \exp\left[i\left(N_{j+}-N_{j-}-Q\right)\phi\right]
  \;=\;
 \int_0^{2\pi}\frac{d\phi}{2\pi}\;\;
   \exp\left[-i\,Q\,\phi \;+\;\sum_{j} z_j\;\left(\lambda_{j+}\;e^{i\phi}
                   \;+\; \lambda_{j-}\;e^{-i\phi}\right)\right]
  \;=\; I_Q(2z)\;.
\end{align}
Parameters $\;\lambda_{j+}\;$ and $\;\lambda_{j-}\;$ in the CE
(\ref{Zce}) are only auxiliary parameters introduced in order to calculate
the mean number and the fluctuations of positively and negatively charged
particles. They are set to one in the final formulas. In Eq.\,(\ref{Zce})
the integral representations of the $\delta$-Kronecker symbol and the
modified Bessel function were used \cite{I}:
\begin{align} \label{IQ}
\delta(n) = \int_0^{2\pi}\frac{d\phi}{2\pi}~ \exp(in\phi)~,\quad I_Q(2z) =
\int_0^{2\pi}\frac{d\phi}{2\pi}\;
   \exp[-i\,Q\,\phi\;+\;2z\,\cos\phi] \;.
\end{align}

The averages of $N_{j+}$ and $N_{j-}$ in both the GCE and
CE can be presented as (in the final expressions one should
put $\lambda_{j\pm}=\exp(\pm\mu/T)$ and $\lambda_{j\pm}=1$ for the
GCE and CE, respectively):
\begin{align}
\label{Njpm} \langle N_{j\pm} \rangle \;=\;
 \left(\lambda_{j\pm}~ \frac{\partial~\ln Z }{\partial\lambda_{j\pm} }
  \right)
 ~=\;
 a_{\pm}~z_j~,
\end{align}
where $a_{\pm}$ in Eq.\,(\ref{Njpm}) is
\begin{align} \label{a+-}
a^{g.c.e.}_{\pm}~=
~\exp\left(\pm\frac{\mu}{T}\right)~,~~~a^{c.e.}_{\pm}~=~\frac{I_{Q\mp
1}(2z)}{I_Q(2z)}~,
\end{align}
for the GCE and CE, respectively.
The average number of $N_{+}$ and $N_{-}$ are equal to:
\begin{align} \label{Npm}
\langle N_{\pm}\rangle \;&=\;
 \langle \sum_j
N_{j\pm}\rangle \;=\; a_{\pm}~\sum_{j}z_{j}~=~a_{\pm}~z~.
\end{align}

 The mean net charge in the GCE  is equal to:
\begin{align}
Q \;=\; \langle N_+\rangle_{g.c.e.} - \langle N_-\rangle_{g.c.e.}
\;=\; 2\sinh\left(\frac{\mu}{T}\right)~z~.\label{Q}
\end{align}
which leads to a simple relation which connects the values of $Q$
and $\mu$
\begin{align}
\exp\left(\frac{\mu}{T}\right) \;=\;
\frac{Q}{2z}+\sqrt{1+\left(\frac{Q}{2z}\right)^2}
 \;\equiv\; y+\sqrt{1+y^2}~,
\end{align}
so that
\begin{align}
\langle N_{\pm}\rangle_{g.c.e.}
 \;=\; z \left(y+\sqrt{1+y^2}\right)^{\pm 1}~,
\end{align}
where $y\equiv Q/2z=\sinh(\mu/T)$.

In the CE an exact charge conservation is imposed on each microscopic
state, so that it is evidently fulfilled also for the average values:
\begin{equation}\label{cons}
\langle N_+ \rangle_{c.e.} ~-~ \langle N_- \rangle_{c.e.} ~ =~
z~\frac{I_{Q-1}(2z)}{I_Q(2z)}~-~z~\frac{I_{Q+1}(2z)}{I_Q(2z)}~=~
 Q~,
\end{equation}
as indeed can be easily seen from the identity $\;
I_{n-1}(x)-I_{n+1}(x)=2nI_n(x)/x \;$ \cite{I}.

The ratios of $\;\langle N_{\pm}\rangle\;$ calculated in the
CE and in the GCE,
\begin{align}\label{Nce-Ngce}
 \frac{\langle N_{\pm} \rangle _{c.e.}}{\langle N_{\pm} \rangle _{g.c.e.}}
 ~=~\frac{I_{Q\mp1}(2z)}{I_Q(2z)}~\cdot~\left(y+\sqrt{1+y^2}\right)^{\mp 1}~,
\end{align}
are shown in Fig.\,1 for $Q=0$ and $Q=2$. There is the strong canonical
suppression effect, $\langle N_{\pm} \rangle _{c.e.}\ll
\langle N_{\pm} \rangle _{g.c.e.}$, for small systems
($z\ll 1$), and the canonical and grand canonical ensembles
become equivalent,
$\langle N_{\pm} \rangle _{c.e.}=\langle N_{\pm} \rangle _{g.c.e.}$,
in the thermodynamic limit $z\rightarrow\infty$.
 One can see that the
CE suppression
effect is reduced for a non-zero net charge of the system
 as compared to a system with
zero net charge. In Fig.\,2 the ratios (\ref{Nce-Ngce}) as
functions of $Q=1,2,...$  are shown at fixed positive values of
$y=Q/2z$ which correspond to the fixed positive net charge number
densities ($Q=0$ corresponds to $y=0$ and  is presented in
Fig.\,1).
Small values of $y$ mean large $z$, e.g. for $y=0.1$ shown in Fig.\,2
one finds `large' $z=5$ at $Q=1$,
so that the system is already close to the
thermodynamic limit. Due to this
the canonical suppression is small and it is the same
for positive and negative particles. The case of large $y$
differs, e.g. for $y=2$
shown in Fig.\,2 the values of $z$ are `small'
at small $Q$: $z=0.25$ at $Q=1$. The canonical suppression
effect becomes strong for
negative particles at small $Q$. However, the canonical suppression
at large $y$ is negligible for the
average value of positive particle number as it should be approximately equal
to $Q$.

 For small systems ($z\ll 1$) using the series expansion \cite{I}
\begin{align} \label{Bessel-1lim}
I_{n}(2z)~=~\frac{z^{n}}{n!}~+~\frac{z^{n+2}}{(n+1)!}~
+~O\left(z^{n+4}\right)~,
\end{align}
one finds for $\;Q=0\;$
\begin{align}\label{ce2}
\langle N_{\pm}\rangle_{c.e.}~\simeq ~z^{2}~~ \ll  ~~ \langle
N_{\pm}\rangle_{g.c.e.}~=~z~,
\end{align}
and for $\;Q\geq 1\;$
\begin{align}\label{NpQ1}
\langle N_{+}\rangle_{c.e.}~&\simeq  ~ Q~,~~~~~ \langle
N_{+}\rangle_{c.e.}~ \simeq~ \langle N_{+}\rangle_{g.c.e.}~;
\\
\langle N_{-}\rangle_{c.e.}~&\simeq ~ \frac{z^{2}}{Q+1}~,
~~~\langle N_{-}\rangle_{c.e.}~\simeq~\frac{Q}{Q+1}~\langle
N_{-}\rangle_{g.c.e.}~.\label{NmQ1}
\end{align}

In the large volume limit ($V\rightarrow \infty$ corresponds also
to $z\rightarrow \infty$) the mean quantities in the CE
and GCE are equal. This result is referred to as an
equivalence of the canonical and grand canonical ensembles.
Using the uniform limit of the modified Bessel function \cite{I}
\begin{align}\label{IQu}
\lim_{n \rightarrow \infty} I_{n}(n x)
 \;=\; \frac{1}{\sqrt{2 \pi n}}\;
       \frac{\exp{(\eta n})}{(1+x^2)^{1/4}}\;
     \left [\; 1 + O\left(\frac{1}{n}\right) \right] \;,
\end{align}
where
\begin{align}\label{eta}
\eta = \sqrt{1+x^2} + \log \frac{x}{1+ \sqrt{1+x^2}}\;,
\end{align}
one can easily find (note that fixed $Q$ at $z\rightarrow
\infty$ means a zero value of the net charge density and $y= 0$):
\begin{align}\label{Npmce}
 \langle N_{\pm} \rangle _{c.e.}~\simeq~
  z\left(y+\sqrt{1+y^2}\right)^{\pm 1}~
 =~\langle N_{\pm} \rangle_{g.c.e.}~.
\end{align}

The total multiplicity of charged 
particles is defined as $N_{ch} = N_++N_-$.
Its average  in the GCE and CE reads: 
\begin{align}\label{Nch-gce} \langle N_{ch} \rangle_{g.c.e.} &\;\equiv\;
\langle\;N_+ + N_-\;\rangle_{g.c.e.} \;=\; \langle N_+\rangle_{g.c.e.} +
\langle N_-\rangle_{g.c.e.} \;= ~2z~\cosh\left(\frac{\mu}{T}\right)\;, \\
\label{Nch-ce} \langle N_{ch} \rangle_{c.e.} &\,\equiv\, \langle\,N_+ +
N_-\,\rangle_{c.e.} \,=\, \langle N_+\rangle_{c.e.} + \langle N_-\rangle_{c.e.}
\,=\, z\left[\frac{I_{Q-1}(2z)}{I_Q(2z)} +
\frac{I_{Q+1}(2z)}{I_Q(2z)}\right]\,.
\end{align}
\begin{figure}[h!]\label{1}
 \vspace{-0.5cm}
 \hspace{-0.1cm}
\epsfig{file=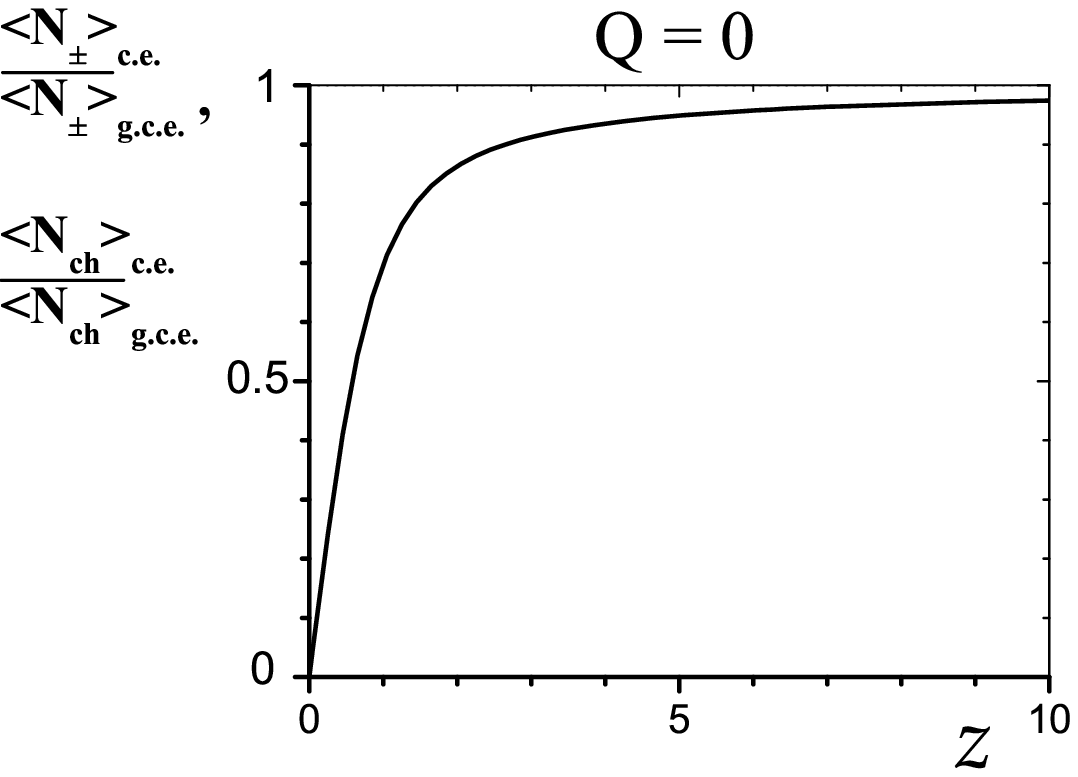,height=7cm,width=9.8cm}
 \hspace{0.1cm}
\epsfig{file=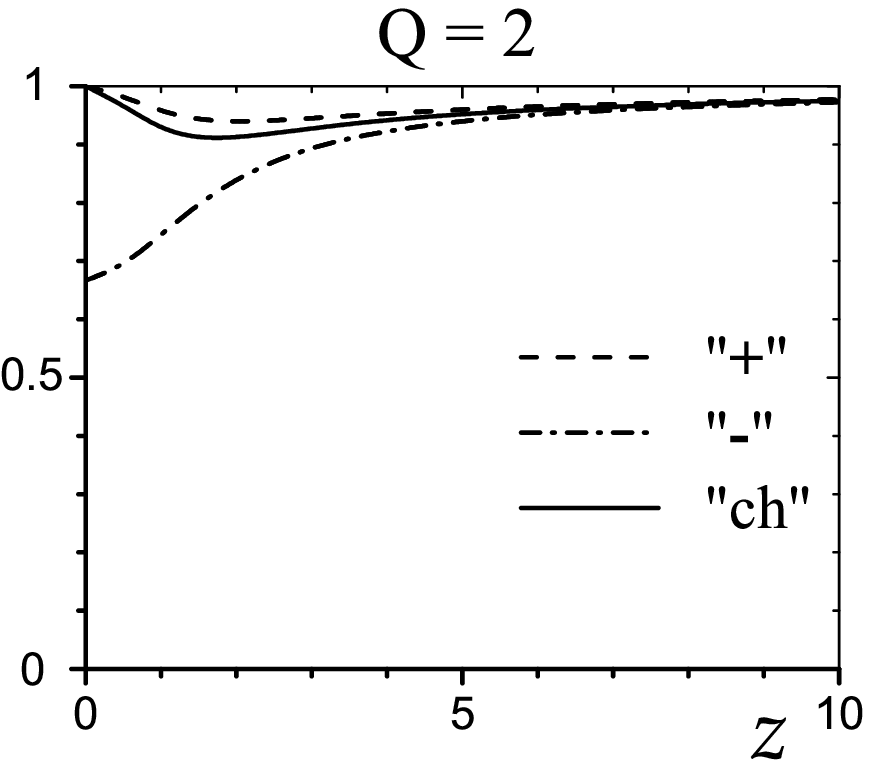,height=7cm,width=7.8cm}
 \vspace{-1cm}
\caption{The ratios of the mean particle numbers in the
CE to those in the GCE as  functions of $z$
for $Q=0$  and $Q=2$.
} %
 \label{NQ02}
\end{figure}
\begin{figure}[h!]\label{2}
 \vspace{0.5cm}
\epsfig{file=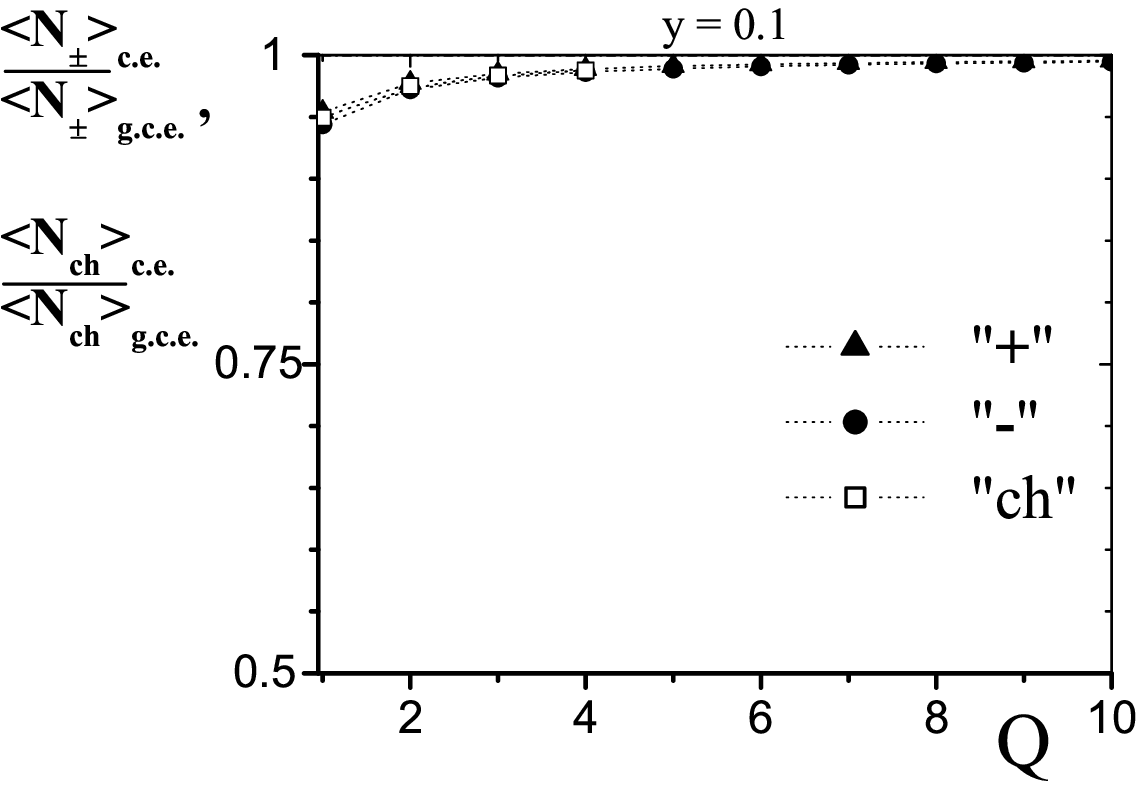,height=7cm,width=9.6cm}
 \hspace{0.05cm}
\epsfig{file=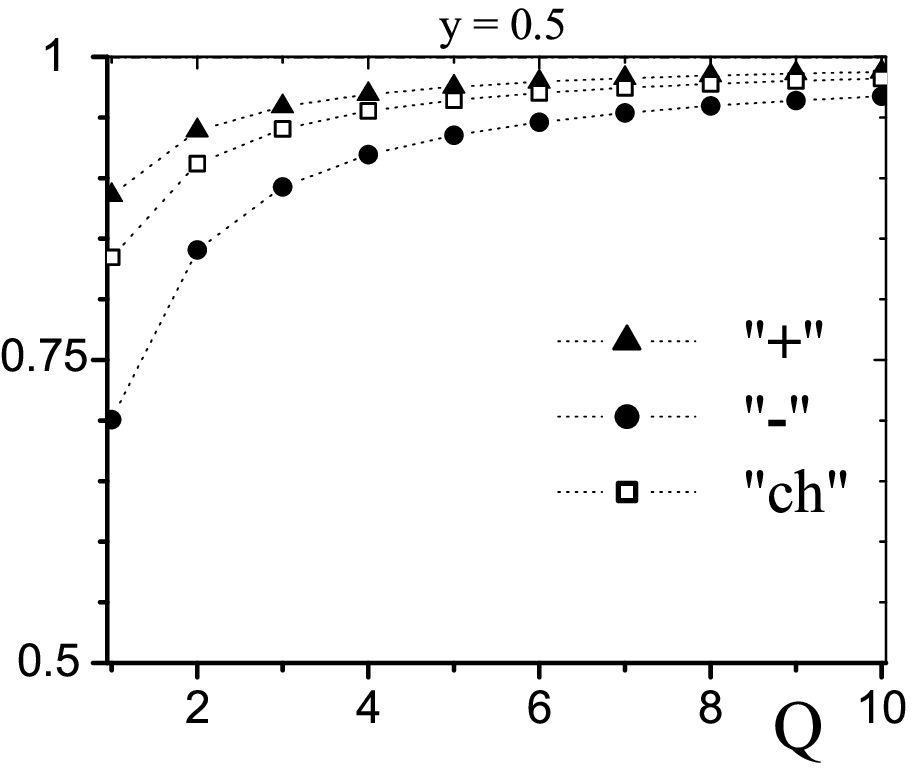,height=7cm,width=7.9cm} \\
 \hspace{1.6cm}
\epsfig{file=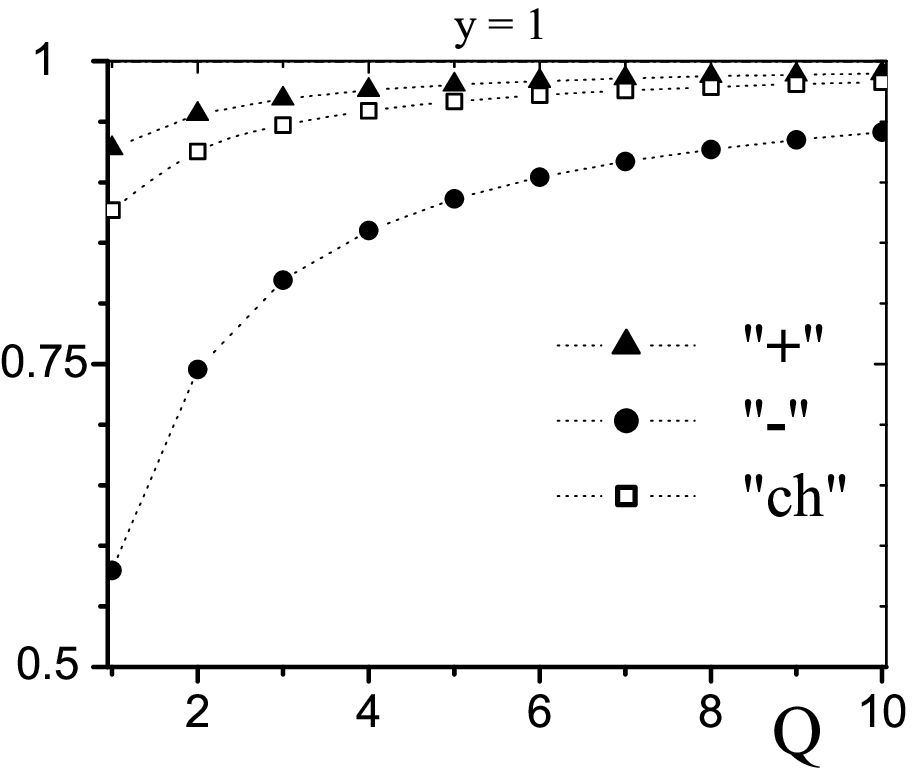,height=7cm,width=8cm}
 \hspace{0.05cm}
\epsfig{file=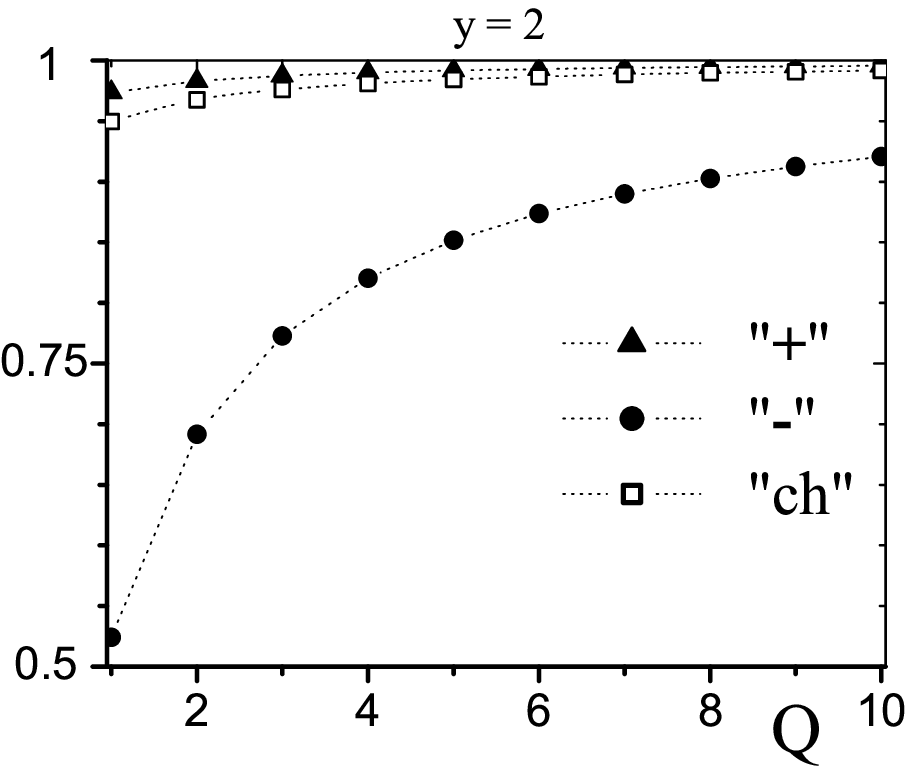,height=7cm,width=7.9cm}
 \vspace{-0.7cm}
\caption{The ratios of the mean particle numbers in the
CE to those in the GCE as  functions
of $Q=1,2,3,...$ for fixed values of $y=Q/2z$. }
\end{figure} %

%
\section{ The scaled variance}
A useful measure of the fluctuations of any variable $\;X\;$ is
the ratio of its variance $\;V(X)\equiv\langle X^2\rangle -\langle
X \rangle ^2\;$ to its mean value $\;\langle X \rangle$, referred
here as the scaled variance:
\begin{align} \label{omega}
 \omega^{X} \;\equiv \; \frac{\langle X^2\rangle - \langle
X\rangle^2}{\langle
X\rangle}~.
\end{align}
Note, that $\;\omega^X=1\;$ for the Poisson distribution. In order
to study the fluctuations of charged particle numbers the second
moments of the multiplicity distribution have to be calculated.
One finds:
\begin{align}\label{Nj2}
\langle N_{j\pm}^{2} \rangle~ &=~ \frac{1}{Z }
  \left[\lambda_{j\pm} \frac{\partial}{\partial \lambda_{j\pm}}
  \left( \lambda_{j\pm} \frac{\partial \; Z}{\partial \lambda_{j\pm}}
  \right) \right]
  = a_{\pm}~z_{j}~+~b_{\pm}~z_{j}^{2}~,\\
\label{Ni+Nj-} \langle N_{j+}N_{j-}\rangle ~&=~ \frac{\lambda_{j+}
\lambda_{j-}}{Z}
    ~\frac{\partial^2\;Z}{\partial \lambda_{j+}\partial \lambda_{j-}}
     ~=~ z_{j}^{2}~,
\end{align}
 where $a_{\pm}$ is given by Eq.\,(\ref{a+-}) and
\begin{align}\label{bpm}
b^{g.c.e.}_{\pm}~=~\exp\left(\pm\frac{2\mu}{T}\right)~=~
\left(a^{g.c.e.}_{\pm}\right)^{2},~~~~~
b^{c.e.}_{\pm}~=~\frac{I_{Q\mp2}(2z)}{I_{Q}(2z)}~,
\end{align}
in the GCE and CE, respectively.
The scaled variances $\omega^{j\,\pm}$ and $\omega^{j\,ch}$ are equal to:
\begin{align}\label{omega-j}
\omega^{j\,\pm}
 \; &\equiv \; \frac{\langle N_{j+}^2\rangle
        \;-\; \langle N_{j+}\rangle^2}
{\langle
N_{j+}\rangle}~=~1\;-\;z_{j}~\left(a_{\pm}~-
~\frac{b_{\pm}}{a_{\pm}}\right)~,\\
\label{omega-ch-j} \omega^{j\,ch}~&\equiv~\frac{\langle \left(N_{j+}+
N_{j-}\right)^2\rangle
        - \langle N_{j+}+ N_{j-}\rangle^2}{\langle
N_{j+}+N_{j-}\rangle}
 ~=~ 1+z_{j}~\left[\frac{b_+  + b_- +2}{a_+ +a_-}-\left(
 a_++a_-\right)\right]~,
\end{align}

\begin{figure}[h!]
 \vspace{-0.7cm}
 \hspace{-0.2cm}
\epsfig{file=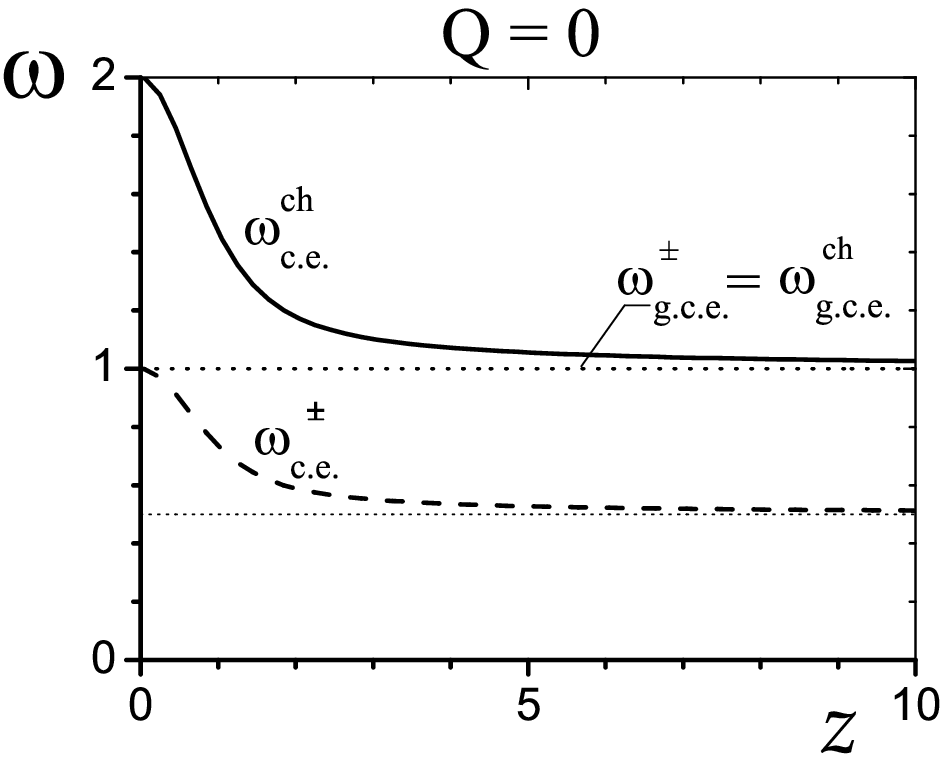,height=7cm,width=9.2cm}
 \hspace{0.4cm}
\epsfig{file=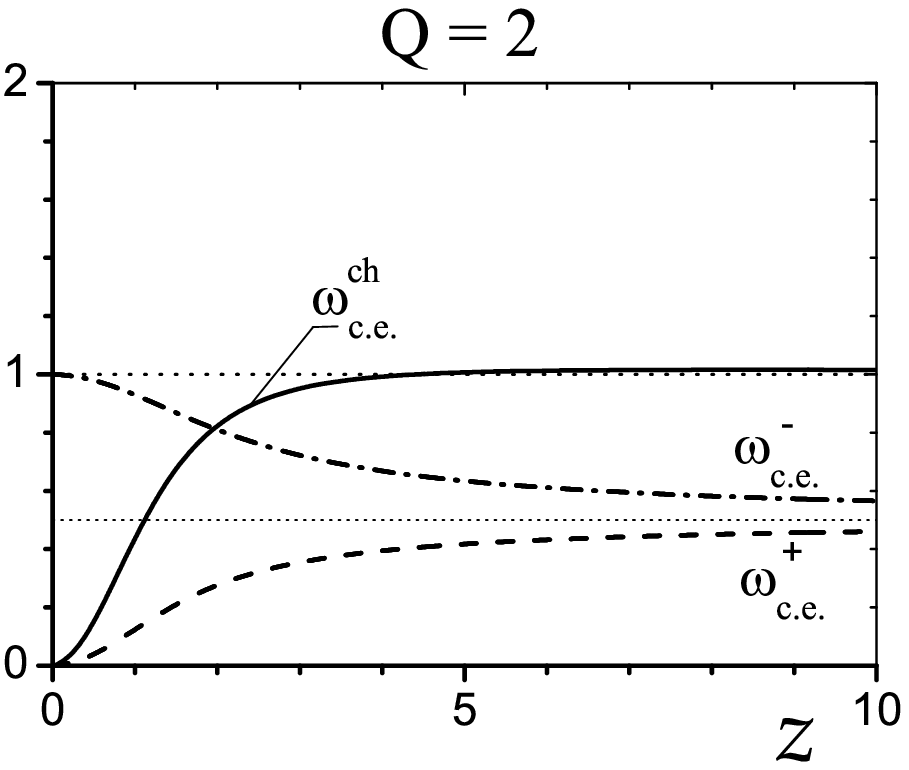,height=7cm,width=8.2cm}
 \vspace{-0.9cm}
 \caption{The scaled variances $\omega_{c.e.}^{\pm}$
 (\ref{omega-ce}) and $\omega_{c.e.}^{ch}$
 (\ref{omega-ch-ce}) as functions of $z$ for fixed values
 of the conserved charge $Q$.}
 \label{wceQ02}
\end{figure}

\begin{figure}[h!]
 \vspace{0.5cm}
 \hspace{-0.2cm}
\epsfig{file=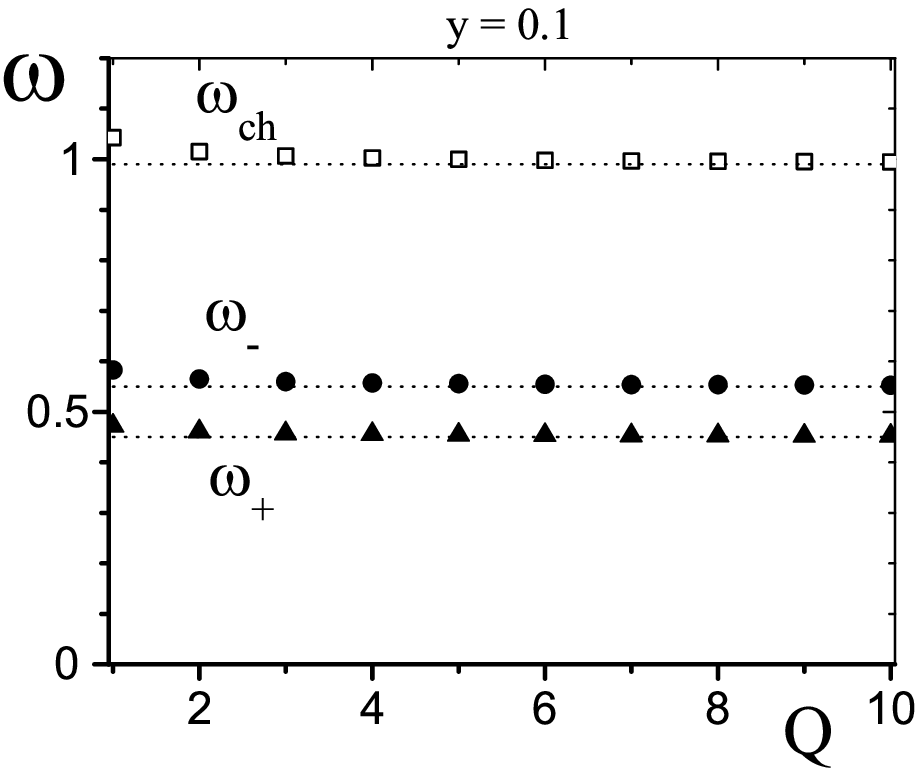,height=7cm,width=8.8cm}
 \hspace{0.3cm}
\epsfig{file=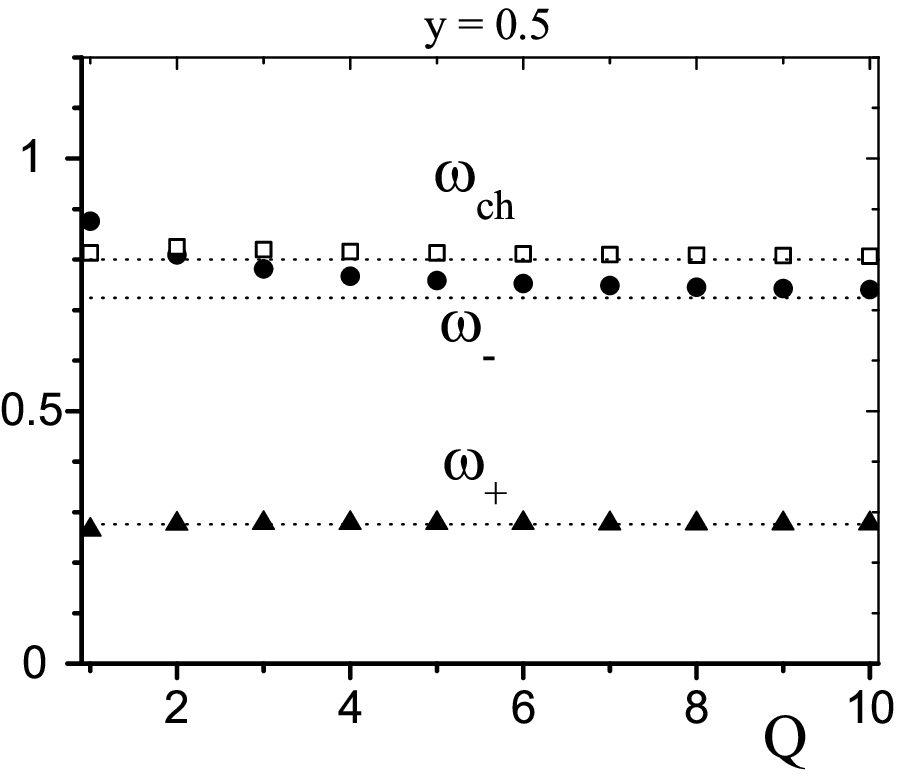,height=7cm,width=8.5cm}
\\
\epsfig{file=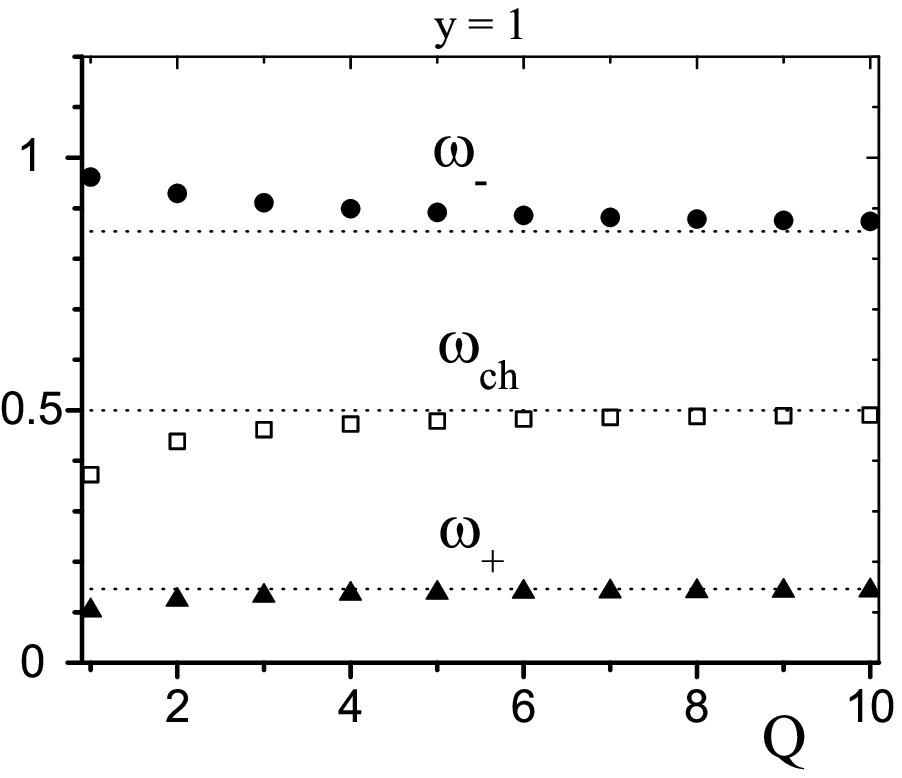,height=7cm,width=8.5cm}
 \hspace{0.3cm}
\epsfig{file=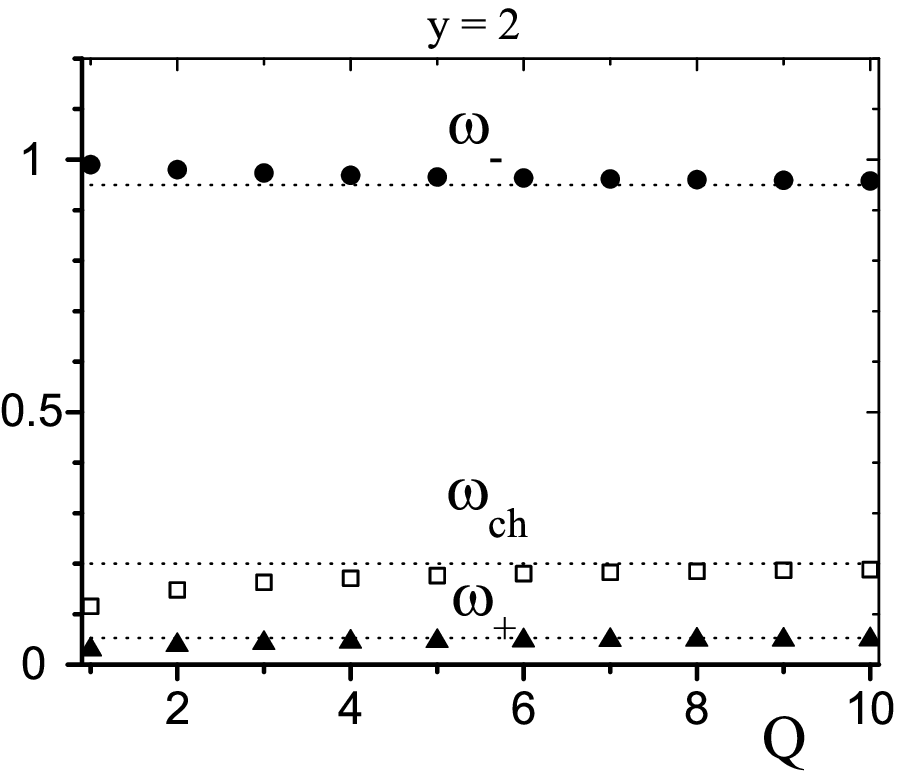,height=7cm,width=8.5cm}
 \vspace{-0.3cm}
 \caption{The scaled variances $\omega_{c.e.}^{\pm}$
 (\ref{omega-ce}) and $\omega_{c.e.}^{ch}$
 (\ref{omega-ch-ce}) as functions of $Q=1,2,3,...$ for fixed values
 of $y=Q/2z$.}
 \label{wcey}
\end{figure}

The Eqs.\,(\ref{omega-j}-\ref{omega-ch-j}) describe the particle number
fluctuations of a given specie ~$j$~. One can establish the general rule
how to calculate the fluctuations of $N_{\pm}=\sum_j N_{j\pm}$ and
$N_{ch}=N_{+}+N_{-}$. To do this one should set
$\lambda_{1\pm}=\lambda_{2\pm}=\cdots=\lambda_{\pm}$ in
Eqs.\,(\ref{Zgce},\ref{Zce}) and differentiate with respect to
$\lambda_{\pm}$ in Eqs.\,(\ref{Njpm},\ref{Nj2}) in order to get $\langle
N^{n}_{\pm} \rangle$ ($n=1,2$). This eventually results in a substitution
of $z$ instead of $z_j$ in all final formulas for the averages and
fluctuations.
One obtains:
\begin{align}\label{omega-gce}
 \omega_{g.c.e.}^{\pm}  ~
&=~\omega_{g.c.e.}^{ch} ~=1~.\\
\omega_{c.e.}^{\pm}
~ & =~
  1 \;-\; z~\left[\,\frac{I_{Q\mp 1}(2z)}{I_Q(2z)}
           \;-\; \frac{I_{Q\mp 2}(2z)}
           {I_{Q\mp 1}(2z)}\,\right]~, \label{omega-ce}\\
   \omega_{c.e.}^{ch}~
  & =\; 1 \;+\; z \left[ \frac{I_{Q-2}(2z) + I_{Q+2}(2z) + 2I_Q(2z)}
                                {I_{Q-1}(2z)+I_{Q+1}(2z)}
       \;-\; \frac{I_{Q-1}(2z)+I_{Q+1}(2z)}{I_Q(2z)} \right]~.
       \label{omega-ch-ce}
\end{align}
The scaled variances $\omega_{c.e}^{\pm}$ and $\omega_{c.e}^{ch}$
calculated with Eqs.\,(\ref{omega-ce}) and (\ref{omega-ch-ce}) are
shown in Fig.\,\ref{wceQ02} for $Q=0,~Q=2$ and in Fig.\,\ref{wcey}
for fixed positive values of $y$. Using  the asymptotic behavior
of the modified Bessel function for $z\rightarrow 0$,
Eq.\,(\ref{Bessel-1lim}), and $z,Q\rightarrow \infty$ with
$y=Q/2z=const$, Eqs.\,(\ref{IQu}-\ref{eta}), 
the limits of the scaled variances can be easily
found, both for a given particle specie ~$j$~
 and for the sum over all particle species:
\\
1). A small system limit $\;z\rightarrow 0\;$ gives for $\;Q=0\;$
\begin{align}
 \omega_{c.e.}^{j\,+}  &\;=\; \omega_{c.e.}^-
 \;\simeq\; 1\;-\;\frac{z_jz}{2}\;,
 &
 \omega_{c.e.}^+  &\;=\; \omega_{c.e.}^-
 \;\simeq\; 1\;-\;\frac{z^2}{2}\;,
 \\
 \omega_{c.e.}^{j\,ch} &\;\simeq\; 1\;+\;\frac{z_j}{z}\;-\; z_jz\;,
 &
 \omega_{c.e.}^{ch} &\;\simeq\; 2\;-\; z^2\;,
\end{align}
and for $\;Q\geq 1\;$
\begin{align}
 \omega_{c.e.}^{j\,+} &\;\cong\; 1\;-\;\frac{z_j}{z}\;+\;\frac{z_jz}{Q(Q+1)}\;,
 &
 \omega_{c.e.}^+ &\;\cong\; \frac{z^2}{Q(Q+1)}\;,
 \\
 \omega_{c.e.}^{j\,-} &\;\cong\; 1 \;-\; \frac{z_jz}{(Q+1)(Q+2)}\;,
 &
 \omega_{c.e.}^- &\;\cong\; 1 \;-\; \frac{z^2}{(Q+1)(Q+2)}\;,
 \\
 \omega_{c.e.}^{j\,ch} &\;\cong\; 1\;-\;\frac{z_j}{z}
  \;+\;\frac{4\,z_jz}{Q(Q+1)}\;,
 &
 \omega_{c.e.}^{ch} &\;\cong\; \frac{4\,z^2}{Q(Q+1)}\;.
\end{align}

2). A large system limit $\;z\rightarrow\infty\;$ gives for fixed
$Q$ (note again that fixed $Q$ in the thermodynamic limit
$z\rightarrow \infty$ means a zero value of the net charge density
and leads, therefore, to $y= 0$)
\begin{align}
 \omega_{c.e.}^{j\,\pm}  &\;\simeq\; 1\;-\;\frac{z_j}{2z}
  \;+\; \frac{z_j}{8z^2} \;\mp\; \frac{Qz_j}{4z^2}\;,
   &
 \omega_{c.e.}^{\pm}  &\;\simeq\; \frac{1}{2} \;+\; \frac{1}{8z}
   \;\mp\; \frac{Q}{4z}\;,
   \\
 \omega_{c.e.}^{j\,ch} &\;\simeq\; 1\;+\; \frac{z_j}{4z^2}\;,
 &
\omega_{c.e.}^{ch} &\;\simeq\; 1\;+\; \frac{1}{4z}\;,
\end{align}
and for fixed $\;Q/2z=y\;$
\begin{align}
 \omega_{c.e.}^{j\,\pm}
 &\;\simeq\; 1\;-\;\frac{z_j}{2z} \;\mp \;
 \frac{z_j}{2z}\frac{y}{\sqrt{1+y^2}}\;,
 &
 \omega_{c.e.}^{\pm}
 &\;\simeq\; \frac{1}{2} \; \mp \;\frac{y}{2\sqrt{1+y^2}}\;,
  \label{omega-plus}
 \\
 \omega_{c.e.}^{j\,ch}
 &\;\simeq\; 1\;-\;\frac{z_j}{z}\frac{y^2}{1+y^2}\;,
 &
 \omega_{c.e.}^{ch}
 &\;\simeq\; \frac{1}{1+y^2}\;.\label{omega-ch}
\end{align}

As  one sees from Eqs.\,3-4 the scaled variances reach very fast
their asymptotic values. In Fig.\,\ref{wceQ02} the scaled variances
for $Q=0$ and $Q=2$ can be compared (for $Q=0$ see details in
Ref.\,\cite{ce-fluc}). One notices that their values at
$z\rightarrow \infty$  are the same, but the behavior at small
values of $z$ is different. Namely, if $Q\geq 1$ the fluctuations
of positively charged particles are very small at small $z$, while
the fluctuations of the negatively charged particles have the
Poisson width. This can be easy understood as for  small volumes
the average number of positive particles is  approximately equal
to $Q$ (see Eq.\,(\ref{NpQ1})) and the fluctuations of $N_+$ are
small. On the other hand, at small $z$ and fixed $Q$ the average
number of
 negatively charged particles is much smaller than $Q$ (see Eq.\,(\ref{NmQ1}))
 and the fluctuations of $N_-$
 are not affected by the conservation law.
Similar physical reasons explain the behavior of the fluctuations
at non-zero charge density in the thermodynamic limit. The Fig.\,4
demonstrates the following features of the asymptotic values of
$\omega_{c.e.}^{+}$ and $\omega_{c.e.}^{-}$ at  $Q\gg 1$. When the
charge density becomes larger ($y$ increases) the
$\omega_{c.e.}^{+}$ decreases and tends to 0 at $y\rightarrow
\infty$, while the $\omega_{c.e.}^{-}$ increases and tends to 1 at
$y\rightarrow \infty$. The physical reasons of this are seen from
Eq.\,(\ref{Npmce}) which at $y\gg 1$ gives: $\langle
N_+\rangle_{c.e.}\simeq z\cdot 2y=Q$ and $\langle
N_-\rangle_{c.e.}\simeq z\cdot(2y)^{-1} =Q\cdot(4y^2)^{-1}\ll Q $.
Therefore, at $y\gg 1$  an exact charge conservation  keeps $N_+$
close to its average value $Q$ and makes  the fluctuations of
$N_+$ in the CE to be small. Under the same conditions,
$\langle N_-\rangle_{c.e.}$ is much smaller than $Q$, so that the
fluctuations of $N_-$  are not affected by the CE suppression
effects and has the Poisson form. These features of the CE are
in a striking difference with those in the GCE. The GCE
scaled variances (\ref{omega-gce}) are equal to 1 for
$N_-$, $N_+$ and $N_{ch}$, and this remains  valid for all
values of the system net charge or net charge density.

\section{ Energy fluctuations}
\begin{figure}[h!]
 \hspace{-0.5cm}
\epsfig{file= 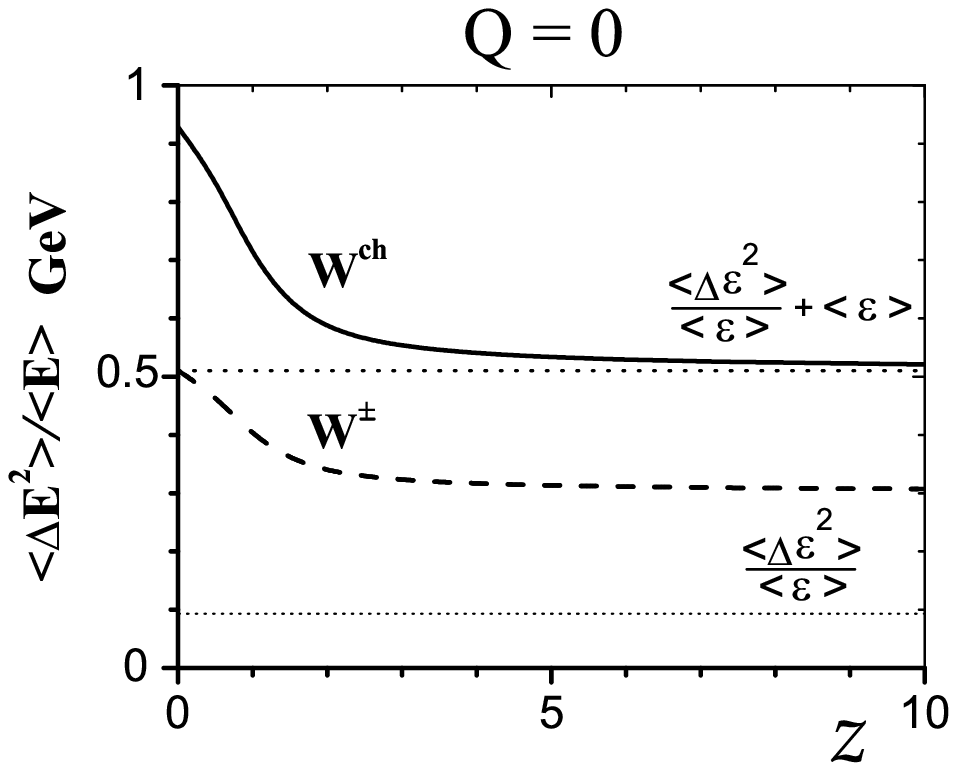,height=7cm,width=9.5cm}
 \hspace{0.1cm}
\epsfig{file=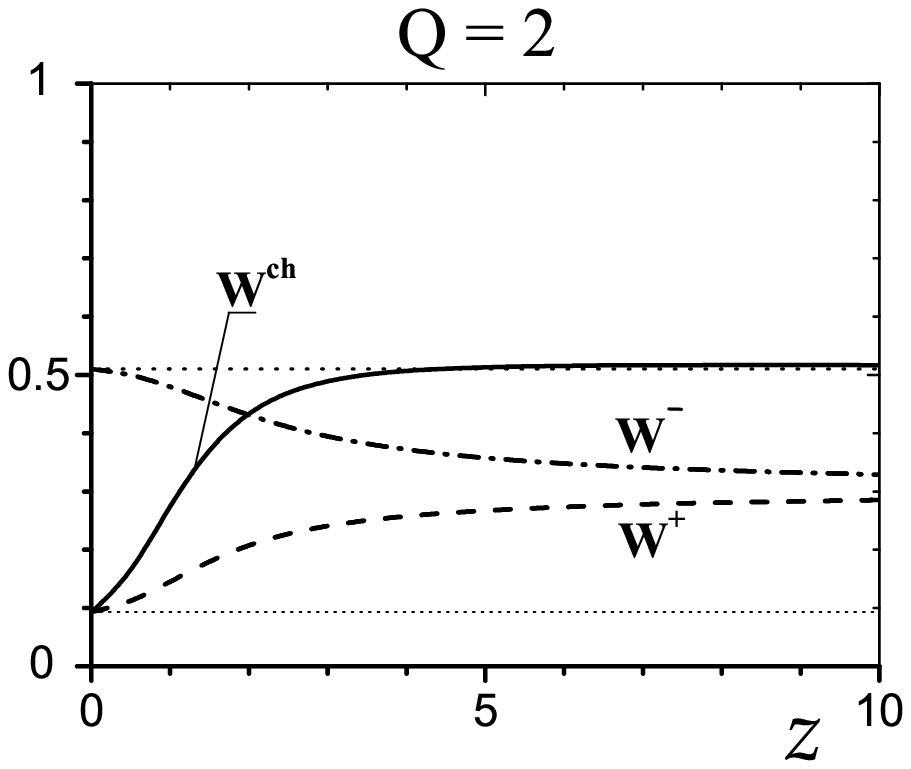,height=7cm,width=8.5cm}
 \vspace{-0.8cm}
\caption{The CE energy fluctuation $W^{+}$, $W^{-}$ and
$W^{ch}$ in the ideal pion gas at temperature $T=120$~MeV.}
 \label{weq02}
\end{figure}

The partition function in the GCE and CE is equal to
$Z\equiv \sum \exp(-E/T)$, where the sum over microstates includes
the  summation (integration) over particle momenta and summation
over number of particles and over different particle species. Each
microscopic  state  has the weight factor $\prod_{j}\exp[(\mu
N_{j+}-\mu N_{j-})/T]$ in the GCE (\ref{Zgce})and
$\delta[Q-\sum_{j}(N_{j+}-N_{j-})]$ in the CE (\ref{Zce}). In
order to calculate the average energy and its fluctuations it is
convenient to rewrite the partition function as $Z =\sum
\exp[\sum_{j}(\beta_{j+}E_{j+}+\beta_{j-}E_{j-})/T]$, where
$\beta_{j+}$ and $\beta_{j-}$ are the auxiliary parameters and
$\beta_{j+}=\beta_{j-}=\beta\equiv 1/T$ in the final formulas. It
then follows:
\begin{align}\label{Ejpm}
\langle E_{j\pm} \rangle~& =~- \frac{1}{Z} \frac{\partial Z}{\partial
\beta_{j\pm}}~=~-~a_{\pm}z_{j}^{'}~\equiv~
\langle \varepsilon_{j}\rangle \langle N_{j\pm}\rangle ,\\
\langle E_{i\pm} E_{j\pm} \rangle~& = ~\frac{1}{Z} \frac{\partial^2
Z}{\partial \beta_{i\pm} \beta_{j\pm}}~ =~
a_{\pm}z_{j}^{''}\delta_{ij}~+~b_{\pm}z_{i}^{'}z_{j}^{'}~,\label{Eijpm} \\
\langle E_{i+} E_{j-} \rangle~& = ~\frac{1}{Z} \frac{\partial^2
Z}{\partial \beta_{i+} \beta_{j-}}~ =~z_{i}^{'}z_{j}^{'}~,\label{Eipjm}
\end{align}
where $z_{j}^{'}=\partial z_{j}/\partial \beta$, $z_{j}^{''}=\partial^{2}
z_{j}/\partial \beta^{2}$, and $z_{j}, a_{\pm}, b_{\pm}$ are given by
Eqs.\,(\ref{z},\ref{a+-},\ref{bpm}), respectively. In Eq.\,(\ref{Ejpm}) we
have introduced the average value of one-particle energy
$\langle \varepsilon_{j}\rangle \equiv -z_{j}^{'}/z_{j}$.
Introducing also $\langle \varepsilon_{j}^{2}\rangle \equiv
z_{j}^{''}/z_{j}$ the energy fluctuations can be then presented
as:
\begin{align}\label{omegaEj}
W^{j\pm}~\equiv~ \frac{\langle E_{j\pm}^{2}\rangle -\langle
E_{j\pm}\rangle^{2}}{\langle E_{j\pm}\rangle}~=~\frac{\langle
\varepsilon_{j}^{2}\rangle -\langle
\varepsilon_{j}\rangle^{2}}{\langle
\varepsilon_{j}\rangle}~+~\langle \varepsilon_{j}\rangle
~\omega^{j\pm}~,
\end{align}
where $\omega^{j\pm}$ is given by Eq.\,(\ref{omega-j}). 
Introducing the total energies
$E_{\pm}\equiv\sum_{j}E_{j\pm}$ and
$E_{ch}\equiv\sum_{j}(E_{j+}+E_{j-})$, one finds:
\begin{align}\label{WEpm}
W^{\pm}~&\equiv~ \frac{\langle E_{\pm}^{2}\rangle -\langle
E_{\pm}\rangle^{2}}{\langle E_{\pm}\rangle}~=~
\frac{\langle \varepsilon^{2}\rangle~ - ~\langle
\varepsilon\rangle^{2}}{\langle \varepsilon\rangle}~+
~\langle \varepsilon\rangle ~\omega^{\pm}~,\\
W^{ch}~&\equiv~ \frac{\langle E_{ch}^{2}\rangle -\langle
E_{ch}\rangle^{2}}{\langle E_{ch}\rangle}~=~
\frac{\langle \varepsilon^{2}\rangle~ - ~\langle
\varepsilon\rangle^{2}}{\langle \varepsilon\rangle}~ +~\langle
\varepsilon\rangle ~\omega^{ch}~,\label{WEch}
\end{align}
where $\langle \varepsilon \rangle\equiv \sum_{j}z_{j}\langle
\varepsilon_{j}\rangle/z$, $\langle \varepsilon^{2}\rangle\equiv
\sum_{j}\langle \varepsilon_{j}^{2}\rangle z_{j}/z$.
The Eqs.\,(\ref{WEpm}-\ref{WEch}) are valid in both the GCE and
CE. The energy fluctuations consist of two terms. The first term  takes 
into account
the fluctuations of one-particle energies, the second one --
the fluctuations of the number
of particles.
Most often 
the
fluctuations of the number of particles are relatively more
important than the fluctuations of one-particle energies. Indeed, the
maximal value of the first term in 
the right hand side of Eqs.\,(\ref{WEpm}-\ref{WEch}) is equal
to $\langle\varepsilon\rangle/3$ for the particles with 
$m/T\rightarrow 0$, and it decreases and goes to zero at
$m/T\rightarrow \infty$. On the other hand, the second term in 
the right hand side of Eqs.\,(\ref{WEpm}-\ref{WEch}) is equal
to $\langle\varepsilon\rangle$ for any system in the GCE.
The value of $(\langle \varepsilon^{2}\rangle - \langle
\varepsilon\rangle^{2})/\langle \varepsilon\rangle$ in
Eqs.\,(\ref{WEpm}-\ref{WEch}) is the same for ``+''and ``-''
particles, and in both the GCE and CE. The values of
$\omega$'s are however different  in the GCE and CE.
Besides, the $\omega^+_{c.e.}$, $\omega^-_{c.e.}$ and
$\omega^{ch}_{c.e.}$  are different from each other for the
non-zero net charge $Q$. Therefore, the scaled variances of the
energy fluctuations are different in the  GCE and CE, and
in the CE the values of $W^{+}$, $W^{-}$ and $W^{ch}$ differ
from each other and depend on the net charge of the system.
 An example of the energy
fluctuations $W^{+}$, $W^{-}$ and $W^{ch}$ for the ideal pion gas with
$Q=0$ and $Q=2$ is presented in Fig.\,\ref{weq02}. One sees that
the dependences of the energy fluctuations $W^{+}$, $W^{-}$ and
$W^{ch}$ on $z$ in the CE resemble those for 
$\omega^+_{c.e.}$, $\omega^-_{c.e.}$ and $\omega^{ch}_{c.e.}$
shown in Fig.\,\ref{wceQ02}.

The upper horizontal dotted line in Fig.\,5 shows the value of
$\langle \varepsilon^{2}\rangle/\langle \varepsilon\rangle$ which
corresponds to the $W^{+}=W^{-}=W^{ch}$ in the GCE. The lower
horizontal dotted line in Fig.\,5 shows the value of $(\langle
\varepsilon^{2}\rangle - \langle \varepsilon \rangle^2)
 /\langle \varepsilon \rangle$ which is the fluctuations
 of one-pion energy.

\section{Single and double charged particles}

 In the following
sections we consider the extension of the CE formalism. First, let us
study the system of particles and antiparticles with charges $\pm 1$ and
$\pm 2$.
%
The CE partition function reads:
\begin{align}\label{ZceMQ}
&Z_{c.e.}(V,T,Q)
 ~=~\sum_{N_{+},N_{-},\widetilde{N}_{+},\widetilde{N}_{-}=0}^{\infty}
 ~\frac{\left(\lambda_{+}z\right)^{N_{+}}}{N_{+}!}~
 \frac{\left(\lambda_{-}z\right)^{N_{-}}}{N_{-}!}~
 \frac{\left(\widetilde{\lambda}_{+}
 \widetilde{z}\right)^{\widetilde{N}_{+}}}{\widetilde{N}_{+}!}
~\frac{\left(\widetilde{\lambda}_{-}\widetilde{z}\right)^{\widetilde{N}_{-}}}
{\widetilde{N}_{-}!}~
\delta\left[\left(N_{+}-N_{-}
+2\widetilde{N}_{+}-2\widetilde{N}_{-}\right)-Q\;\right]
\nonumber
\\
 &\;=\;
 \int_0^{2\pi}\frac{d\phi}{2\pi}\;
 \exp\left[-i\,Q\,\phi \;+\; z~\left(\lambda_{+}\,e^{i\phi}
                   \;+\; \lambda_{-}\,e^{-i\phi}\right)
                   \;+\;
           \widetilde{z}\,\left(\widetilde{\lambda}_{+}\,e^{i2\phi}
                   \;+\;
           \widetilde{\lambda}_{-}\,e^{-i2\phi}\right)\right]
   \;=\; \sum_{k=-\infty}^{\infty}\;\;I_{Q-2k}(2z)\;I_k(2\widetilde{z})
        \;,
\end{align}
where we have used the relation
$\exp\left[x\left(t+\frac{1}{t}\right)\right]=\sum_{k=-\infty}^{\infty}t^k
I_k(2x)\;$.
%
The  $z$ and $\tilde{z}$ in Eq.\,(\ref{ZceMQ}) are the one-particle
partition functions for charges $\pm 1$ and $\pm 2$, respectively.
In terms of variables $c_{m\pm}$ ($m=1,2,4$)
\begin{align}\label{cnpm}
c_{m\pm}
 \;=\; \frac{\sum_{k=-\infty}^{\infty}\;\;
             I_{Q\mp m-2k}(2z)\;I_k(2\widetilde{z})}
            {\sum_{k=-\infty}^{\infty}\;\;
             I_{Q-2k}(2z)\;I_k(2\widetilde{z})}\;,
\end{align}
one finds:
\begin{align} \label{N1N2}
 \langle N_{\pm}\rangle_{c.e.} &\;=\; c_{1\pm}\;z\;, ~~&
 \langle \widetilde{N}_{\pm}\rangle_{c.e.} &\;=\; c_{2\pm}\;\widetilde{z}\;
 ,
 \\
 \langle N_{\pm}^2\rangle_{c.e.}
 &\;=\; c_{1\pm}\;z \;+\; c_{2\pm}\;z^2,
 ~~&
 \langle \widetilde{N}_{\pm}^2\rangle_{c.e.}
 &\;=\; c_{2\pm}\;\widetilde{z} \;+\; c_{4\pm}\;\widetilde{z}^2\;
 . \label{N1N2a}
\end{align}
From Eqs.\,(\ref{N1N2}-\ref{N1N2a}) it follows:
\begin{align}
\omega_{1\,c.e.}^{\pm}
 & \;\equiv \; \frac{\langle N_{\pm}^2\rangle_{c.e.}
        \;-\; \langle N_{\pm}\rangle_{c.e.}^2}
   {\langle N_{\pm}\rangle}_{c.e.}~=
   ~1~-~z~\left(c_{1\pm}~-~\frac{c_{2\pm}}{c_{1\pm}}\right)~, \label{omega1}
 \\
\omega_{2\,c.e.}^{\pm}
 & \;\equiv \; \frac{\langle \widetilde{N}_{\pm}^2\rangle_{c.e.}
        \;-\; \langle \widetilde{N}_{\pm}\rangle^2_{c.e.}}
   {\langle \widetilde{N}_{\pm}\rangle_{c.e.}}~=
   ~1~-~\widetilde{z}~
   \left(c_{2\pm}~-~\frac{c_{4\pm}}{c_{2\pm}}\right)~,\label{omega2}
 \\
 \omega_{c.e.}^{\pm}
 & \;\equiv \; \frac{\langle\left( N_{\pm}+
 \widetilde{N}_{\pm}\right)^2\rangle_{c.e.}
        \;-\; \langle N_{\pm}+ \widetilde{N}_{\pm}\rangle^2_{c.e.}}
   {\langle N_{\pm}+ \widetilde{N}_{\pm}\rangle_{c.e.}} \label{omega12}
 ~=~1~+~\frac{z^2c_{2\pm}
      \;+\;\widetilde{z}^2c_{4\pm}
        \;+\; 2z\widetilde{z} c_{3\pm}}
      {z c_{1\pm}\;+\;\widetilde{z} c_{2\pm}}
  \;-\; (z c_{1\pm}\;+\;\widetilde{z} c_{2\pm})~.
\end{align}
To illustrate the specific features of the considered system we
present the CE results in the case $\;Q=0\;$.  As always, all
$\omega$'s are equal to 1 in the GCE. For $\langle Q\rangle
_{g.c.e.}=0$ one also has $\langle N_{\pm}\rangle_{g.c.e.}=z$ and
$\langle \widetilde{N}_{\pm}\rangle_{g.c.e.}=\widetilde{z}$.
To calculate Eqs.\,(\ref{N1N2}--\ref{omega12}) for finite values of
$z$ and $\widetilde{z}$ one can effectively use Eq.\,(\ref{cnpm}).
In the thermodynamic limit $V\rightarrow \infty$ both
$z\rightarrow \infty$ and $\widetilde{z} \rightarrow\infty$. In
this case it is convenient to return  to the integral over $\phi$
representation in Eq.\,(\ref{ZceMQ}) and use it also for the
derivatives of the CE partition function with respect to
$\lambda_{\pm}$ and $\widetilde{\lambda}_{\pm}$. Using the saddle
point method to calculate the $\phi$-integrals one finds then for
$z,\widetilde{z}\gg1$:
\begin{align}
\langle N_{\pm}\rangle_{c.e.}
 & ~
\simeq ~z\left[1 - \frac{1}{4(z+4\widetilde{z})}\right]~,~~
\langle \widetilde{N}_{\pm}\rangle_{c.e.}~=~ \widetilde{z}\left(1
- \frac{1}{z+4\widetilde{z}}\right)~;
\\
\omega_{1\,c.e.}^{\pm} &\; \simeq\; 1-\frac{z}{2(z+4\widetilde{z})}\;,~~
\omega_{2\,c.e.}^{\pm} \; \simeq \;
1-\frac{2\widetilde{z}}{z+4\widetilde{z}}\;,~~
\omega_{c.e.}^{\pm} \; \simeq\;
1-\frac{(z+2\widetilde{z})^2}{2(z
+\widetilde{z})(z+4\widetilde{z})}\;.\label{W12}
\end{align}
From the above formulas one finds that $ \omega_{c.e.}^{\pm} \simeq 0.5$
if $\widetilde{z}/z$ is either much smaller or much larger than 1.
The scaled variance $\omega_{c.e.}^{\pm}$ has a maximum at
$z=2\widetilde{z}\;$. At this point one finds $\omega_{c.e.}^{\pm}=5/9$,
$\omega_{1\,c.e.}^{\pm}=5/6$ and $\omega_{2\,c.e.}^{\pm}=2/3$.
\section{Quantum statistics effects}

It is instructive to apply a different technique \cite{steph} to
calculate the fluctuations of the thermodynamical quantities with
the exact conservation laws imposed. This method allows to find
the values of the CE fluctuations in the
thermodynamic limit and include the effects of  quantum
statistics.

The ideal quantum gas of the identical Bose or Fermi particles and 
antiparticles can be
characterized by the occupation numbers 
$n_{p}^{\pm}$ 
of the one-particle states
labeled by momenta p. The GCE average values and fluctuations are
\cite{lan}:
 \eq{
 \langle n_p^{\pm} \rangle_{g.c.e.}
 ~& = ~\frac {1}
 {\exp \left[\left(\sqrt{p^{2}+m^{2}}~\mp~ \mu\right) / T\right]
 ~-~ \gamma}~, \label{np-aver}\\
\langle\Delta n_p^{\pm 2}\rangle_{g.c.e.} ~& \equiv~ \langle
\left(n_{p}^{\pm}\right)^2\rangle_{g.c.e.}~-~
\langle n_{p}^{\pm}\rangle^{2}_{g.c.e.}~=~
\langle n_p^{\pm} \rangle_{g.c.e.} \left(1 + \gamma \langle n_p^{\pm}
\rangle_{g.c.e.}\right)
~ \equiv~
  v^{\pm 2}_p~,\label{np-fluc}
  }
where $\gamma$ is equal to $+1$ and $-1$ for Bose and Fermi
statistics, respectively ($\gamma=0$ corresponds to the Boltzmann
approximation).
 These expressions are microscopic  in a sense that they describe
 the average values and fluctuations of a single mode with momentum $p$.
 However, the average values 
 of all macroscopic quantities of the system can be determined
  through the average occupation numbers 
  of these single modes. The fluctuations
  of the macroscopic observables can be written
  in terms of the microscopic correlator $\langle \Delta n_p^\alpha \Delta
  n_k^\beta \rangle_{g.c.e.}$, where $\alpha,\beta$ are $+$ and(or) $-$.
  This correlator can be presented as:
  \eq{
  \langle \Delta n_p^\alpha \Delta n_k^\beta \rangle_{g.c.e.}~ =~
  v_p^{\alpha2}~
  \delta_{p\,k}~\delta_{\alpha \beta}~,  \label{correlator1}
  }
   due to the statistical independence of different
  quantum levels and different charge states in the GCE.
 The variances of the 
 total number of (negatively) positively charged  particles
  $N_{\alpha} = \sum_p n_p^\alpha$ are equal to:
 \eq{
 \langle \Delta N_{\alpha}^2\rangle_{g.c.e.}\,  \equiv\,
 \langle N_{\alpha}^{2} \rangle_{g.c.e.} - \langle N_{\alpha}
 \rangle^2_{g.c.e.}\,=\, \sum_{p,k} \left(\langle n_p^\alpha n_k^\alpha
 \rangle_{g.c.e.} -
  \langle n_p^\alpha \rangle_{g.c.e.} \langle n_k^\alpha \rangle_{g.c.e.}
  \right)\,
 =\, \sum_{p,k} \langle \Delta n_{p}^{\alpha} \Delta_k^\alpha
 \rangle_{g.c.e.}\, = \, \sum_p v_p^{\alpha 2}~.\nonumber
}
 We have assumed above that the quantum $p$-levels
 are non-degenerate. In fact each  level should be
 further specified by the projection of a particle
 spin. Thus, each $p$-level splits into $g=2j+1$
 sub-levels. It will be assumed that the $p$-summation includes all
 these sub-levels too.
 The degeneracy factor enters explicitly when one substitutes, in the
 thermodynamic limit,
the summation over discrete levels by the integration:
\eq{\sum_{p}~...~=~\frac{gV}{2\pi^{2}}\int_{0}^{\infty}p^{2}dp~...~.
\nonumber}
The scaled variance $\omega^{\alpha}_{g.c.e.}$ in the thermodynamical limit
$V\rightarrow\infty$ reads:
 \eq{\omega^{\alpha}_{g.c.e.} ~\equiv~\frac{\langle N_{\alpha}^2
 \rangle_{g.c.e.}~-
 ~\langle N_{\alpha}\rangle^{2}_{g.c.e.}}
 {\langle N_{\alpha} \rangle_{g.c.e.}}~=~ \frac{\sum_{p,k} \langle
 \Delta n_{p}^\alpha \Delta n_{k}^\alpha \rangle_{g.c.e.}}
 {\sum_p \langle n^\alpha_p\rangle_{g.c.e.}}~ = ~
 \frac{\sum_{p}v_p^{\alpha 2}}{\sum_{p}\langle n_p^\alpha\rangle_{g.c.e.}}
 ~\simeq~
\frac{\int_{0}^{\infty}p^{2}dp~v_p^{\alpha
2}}{\int_{0}^{\infty}p^{2}dp~\langle n_p^\alpha\rangle_{g.c.e.}
}~.\label{omega-alpha}
 }
The Eq.\,(\ref{omega-alpha})
corresponds to the particle number
fluctuations in the GCE. To illustrate the role of quantum
statistics let us consider the case of $\mu=0$, i.e. $\langle
Q\rangle_{g.c.e.}=0$, where $Q\equiv\sum_{p,\alpha}
q^{\alpha}n_{p}^{\alpha}$. In what follows we assume $q^+=1$
and $q^-=-1$, however, the formulas below are valid for any values
of $q^+$ and $q^-=-q^+$. From Eqs.\,(\ref{np-fluc}) and (\ref{omega-alpha})
 one finds
$\omega^{\pm Boltz}_{g.c.e.}=1$ in the Boltzmann limit ($\gamma=0$),
$\omega_{g.c.e.}^{\pm Bose}>1$ for the Bose particles ($\gamma=1$) and
$\omega_{g.c.e.}^{\pm Fermi}<1$ for the Fermi particles ($\gamma=-1$).
The strongest quantum effects correspond to $m/T\rightarrow 0$:
\eq{
\omega_{g.c.e.}^{\pm Boltz}~=~1,~~~
\omega_{g.c.e.}^{\pm Bose}~=~
\frac{\pi^{2}}{6\,\zeta(3)}~ \simeq ~1.368~,~~~
\omega_{g.c.e.}^{\pm Fermi}~=~
\frac{\pi^{2}}{9\,\zeta(3)}~ \simeq ~0.912~. \label{omegaBFgce} }
The scaled variance
$\;\omega^{ch}_{g.c.e.}\;$ for all charged particles can be
easily obtained from
(\ref{omega-alpha}) by replacing
$\;\sum_p\rightarrow\sum_{p\,\alpha}\;$, and one finds:
\eq{
\omega_{g.c.e.}^{ch ~Boltz}~=~\omega_{g.c.e.}^{\pm Boltz}~,~~~
\omega_{g.c.e.}^{ch ~Bose}~=~\omega_{g.c.e.}^{\pm Bose}~,~~~
\omega_{g.c.e.}^{ch ~Fermi}~=~\omega_{g.c.e.}^{\pm Fermi}~.\
\label{omegaBFgcech}
}

  The formula for the microscopic correlator is modified if we
  impose the exact charge conservation  in our equilibrated system.
   For this purpose we introduce the equilibrium
  probability distribution $W(n_p^{\alpha})$ of the occupation numbers.
As a first step we assume that each $n_{p}^{\alpha}$ fluctuates
independently according
 to the Gauss distribution law with 
 the average value $\langle n_p^{\alpha}\rangle_{g.c.e.}$ (\ref{np-aver}) and
 the mean square deviation $v_p^{\alpha 2}$ (\ref{np-fluc}):
 \eq{
   W(n_p^\alpha) ~\propto ~ \prod_{p,\alpha}
   \exp{\left[-~ \frac{\left(\Delta n_p^{\alpha}\right)^2}{2v_p^{\alpha 2}}
    \right]}~.
\label{gauss}
}
To justify this assumption (see Ref.\,\cite{steph})
one can consider the
  sum of $n^{\alpha}_p$ in small momentum volume $\left(\Delta
  p\right)^3$ with the center at $p$. At fixed $\left(\Delta
  p\right)^3$ and $V\rightarrow \infty$ the average number of particles
  inside $\left(\Delta p\right)^3$ becomes large.
  Each particle configuration inside $\left(\Delta p\right)^3$
   consists of $\left(\Delta p\right)^3 \cdot V/(2\pi)^3\gg1$ statistically
  independent terms, each
with the average value $\langle n^{\alpha}_{p}\rangle_{g.c.e.}$
(\ref{np-aver}) and the scaled variance
$v^{\alpha 2}_{p}$ (\ref{np-fluc}). From the central limit
  theorem it then follows that
  the probability distribution for the fluctuations
  inside $\left(\Delta p\right)^3$ should be Gaussian.
In fact, we always convolve $n^{\alpha}_p$ with some smooth
  function  of $p$, so instead of writing the Gaussian
  distribution for the sum of $n^{\alpha}_p$ in $\left(\Delta p\right)^3$
we can use it directly for $n_{p}^{\alpha}$.

   The average value of the conserved charge
   $Q=\sum_{p,\alpha}q^{\alpha} n_p^{\alpha}$
   is regulated in the GCE by the chemical potential
   $\mu$.  If we impose an exact charge conservation,
   $\Delta Q=\sum_{p,\alpha} q^{\alpha} \Delta n_p^{\alpha}=0$, the
   distribution (\ref{gauss}) will be modified as:
\begin{align}\label{gauss-Q}
   W(n_p^\alpha) ~\propto ~\prod_{p,\alpha}
   \exp\left[-~ \frac{\left(\Delta n_p^{\alpha}\right)^2}{2v_p^{\alpha 2}}
      \right]~
    \delta\left(\sum_{p,\alpha} q^{\alpha} \Delta n_p^{\alpha}
    \right)~
     \propto ~  \int_{-\infty}^{\infty} d \lambda~\prod_{p,\alpha}
  \exp\left[-~ \frac{\left(\Delta n_p^{\alpha}\right)^2}{2v_p^{\alpha 2}}
+ i \lambda~q^{\alpha} \Delta n_p^{\alpha} \right]~ .
    \end{align}
 It is convenient to generalize distribution (\ref{gauss-Q})
 to $W(n_p^\alpha,\lambda)$ using
 further the integration along imaginary axis in the
 $\lambda$-plane.
After completing squares one gets:
   \eq{ W(n^{\alpha}_p, \lambda)~ \propto ~  \prod_{p, \alpha}
    \exp\left[
    -~\frac{\left(\Delta n_p^{\alpha} -
    \lambda v_p^{\alpha 2}q^{\alpha}\right)^2}
    {2v_p^{\alpha 2}}+
    \frac{\lambda^2}{2}~ v_p^{\alpha 2}q^{\alpha 2}\right]~,
    }
and the average values are now calculated as:
\eq{\langle ... \rangle~=~\frac{\int_{-i\infty}^{i\infty}d\lambda
\int_{-\infty}^{\infty}\prod_{p,\alpha} dn_p^{\alpha}~... ~W(n^{\alpha}_p,
\lambda)}{\int_{-i\infty}^{i\infty}d\lambda
\int_{-\infty}^{\infty}\prod_{p,\alpha} dn_p^{\alpha}~W(n^{\alpha}_p,
\lambda)}~. \label{average} }
The Eq.\,(\ref{average}) has the meaning of the
    CE averaging in the thermodynamic limit $V\rightarrow\infty$.
    One easily finds
    \eq{
    \langle(\Delta n_p^\alpha -
    v_p^{\alpha 2}\lambda q^{\alpha})
    (\Delta n_k^{\beta} - v_k^{\beta 2}\lambda
    q^{\beta})   \rangle ~=~
    \delta_{p\,k}~\delta_{\alpha \beta}~ v_p^{\alpha 2}~,~~~
   &\langle \lambda^2 \rangle = - \left( \sum_{p,
    \alpha} v_p^{\alpha 2} q^{\alpha 2} \right)^{-1}~,
  ~~~
  \langle (\Delta n_p^{\alpha} - v_p^{\alpha 2}\lambda q^{\alpha}) \lambda
  \rangle ~=~ 0~,\nonumber
  }
 so that it follows:
  \begin{align}
   \langle \Delta n^{\alpha}_p \Delta n^{\beta}_k \rangle ~& =~
   \delta_{p\,k}~ \delta_{\alpha \beta}~
   v_p^{\alpha 2}~ - v_p^{\alpha 2} q^{\alpha}~
   v_k^{\beta 2} q^{\beta}~ \langle \lambda^2 \rangle +
   \langle \Delta n_p^{\alpha} \lambda \rangle~ v_k^{\beta 2}
   q^{\beta} + \langle \Delta n_k^{\beta} \lambda \rangle~ v_p^{\alpha 2}
   q^{\alpha} \label{corr-ch}
   \\
   &=~ \delta_{p\,k}~ \delta_{\alpha \beta}~
   v_p^{\alpha 2}~ +~ v_p^{\alpha 2} q^{\alpha}~
   v_k^{\beta 2} q^{\beta} ~\langle \lambda^2 \rangle~
   = ~\delta_{p\,k}~ \delta_{\alpha \beta}~ v_p^{\alpha 2} -
   \frac{v^{\alpha 2}_p q^{\alpha}~v^{\beta 2}_k  q^{\beta}}
   {\sum_{p,\alpha} v^{\alpha 2}_p q^{\alpha 2}~.} \nonumber
  \end{align}
   By means of Eq.\,(\ref{corr-ch}) we obtain:
   \eq{ \omega^{\alpha}_{c.e.}~\equiv~\frac{
    \langle N_{\alpha}^2 \rangle ~-~ \langle N_{\alpha} \rangle^2}
    {\langle N_{\alpha} \rangle} ~=~\frac{
    \sum_p v_{p}^{\alpha 2}}{\sum_{p}\langle n_{p}^{\alpha}\rangle_{g.c.e.}}
    ~-~
    \frac{\left(\sum_p v_{p}^{\alpha 2}q^{\alpha}\right)^{2}}
    {\sum_{p}\langle n_{p}^{\alpha}\rangle_{g.c.e.}
    ~\sum_{p,\alpha} v_{p}^{\alpha
    2}q^{\alpha 2}
    } ~. \label{omega-alpha-ce}
   }
 The Eq.\,(\ref{np-aver}) leads to
  $v_{p}^{\alpha 2}=\langle
n_{p}^{\alpha}\rangle_{g.c.e.}$
in the Boltzmann approximation, so that from Eq.\,(\ref{omega-alpha-ce})
one finds
($y\equiv Q/2z=\sinh(\mu/T)$):
\eq{\omega^{\alpha}_{c.e.}~=~1~-~\frac{\exp(\alpha
\mu/T)}{\exp(\mu/T)~+~\exp(-\mu/T)}~=~ \frac{1}{2}~-~\alpha
~\frac{y}{2\sqrt{1+y^{2}}}~,\label{omega-alpha1}
}
which coincides with Eq.\,(\ref{omega-plus}). Formula
$\;\omega^{ch}_{c.e.}\;$ can be obtained from
(\ref{omega-alpha-ce}) after replacing
$\;\sum_p\rightarrow\sum_{p,\alpha}\;$, and it is the same as
Eq.\,(\ref{omega-ch}).
%
%
At $\mu=0$  from Eq.\,(\ref{omega-alpha-ce})
we find the CE scaled variances:
 \eq{ \omega_{c.e.}^{\pm Boltz} & \;=\;\frac{1}{2}\,,&
      \omega_{c.e.}^{\pm Bose}~ & =~ 
      \frac{\pi^{2}}{12~\zeta(3)}~ \simeq ~0.684\,,&
      \omega_{c.e.}^{\pm Fermi}~ & =
      ~ \frac{\pi^{2}}{18~\zeta(3)}~ \simeq ~0.456\,, \label{omegaBFce}
      \\
\omega_{c.e.}^{ch ~Boltz}~& =~2\,\omega_{g.c.e.}^{\pm Boltz}\,,&
\omega_{c.e.}^{ch ~Bose}~ &=~2\,\omega_{c.e.}^{\pm Bose}\,,&
\omega_{c.e.}^{ch ~Fermi}~ &=~2\,\omega_{c.e.}^{\pm Fermi}\,.
\label{omegaBFcech} }

As seen from Eqs.\,(\ref{omegaBFgce},\ref{omegaBFce}) the scaled
 variance of (negative) positive  particles
 with Bose or Fermi statistics in the CE
 is as half as large as the corresponding scaled variances in the GCE.
 Therefore, the CE suppression of the particle number fluctuations in the
 thermodynamic limit works at $\mu=0$ in the quantum systems similar
 to that in the Boltzmann case. This result can be rephrased in another way:
 the Bose enhancement and Fermi suppression of the GCE fluctuations
 remain the same in the CE  for the $\omega_{c.e.}^{\pm}$ at
$\mu=0$ in the  thermodynamic limit.
 The Eq.\,(\ref{omegaBFcech}) demonstrates that the
  scaled variances of all charged particles in the CE
 for any statistics are by a factor of 2 larger than the corresponding
 scaled variances for (negative) positive  particles,
whereas in the GCE
   these scaled variances presented by Eq.\,(\ref{omegaBFgcech})
   are equal to each other.

  Comparing Eq.\,(\ref{corr-ch}) and Eq.\,(\ref{correlator1}) one notices the
  changes of the microscopic correlator due to an exact charge
  conservation. Namely, in the CE 
the fluctuations of each mode is reduced, i.e.
the $\langle \left(\Delta n_{p}^{\alpha }\right)^2\rangle$ calculated 
 from Eq.\,(\ref{corr-ch}) is smaller than that in
 Eq.\,(\ref{np-fluc}), and the anticorrelations 
  between different modes $p\neq k$ 
  and the same charge states $\alpha=\beta$
  appear. 
These two changes of the microscopic correlator result in a suppression of
the CE scaled variances $\omega^{\alpha}_{c.e.}$  in a comparison with
the GCE ones $\omega^{\alpha}_{g.c.e.}$ (compare Eq.\,(\ref{omega-alpha-ce})
and Eq.\,(\ref{omega-alpha})). Therefore, 
the fluctuations of both $N_-$ and $N_+$
are always suppressed in the CE. 
As we have seen in the previous sections the behavior
  of $N_{ch}$ fluctuations in the CE can be more complicated. This is because
  of the correlations of different modes $p\neq k$  
for the different charge states $\alpha= -\beta$ (i.e. 
the second term in the right
hand side of Eq.\,(\ref{corr-ch}) is positive for $\alpha= -\beta$).

 The exact charge conservation
should also lead to the canonical suppression of $\langle
n_{p}^{\alpha}\rangle$, and this should result in the canonical
suppression effects for $ \langle N_{\alpha}\rangle$. They are,
however, absent in the present formulation, so that formula
(\ref{corr-ch}) for the microscopic correlator is not enough to
calculate $\langle N^2_{\alpha} \rangle$ and $\langle N_{\alpha}
\rangle^{2}$ separately with an accuracy corresponding to the
effects of the canonical suppression. Nevertheless,
  it does allow us to calculate their difference
  $\langle \left(\Delta N_{\alpha}\right)^2  \rangle$
  with the effects of the CE correctly included. This means that
  the canonical suppression
   effects in the occupation numbers $ \langle n_{p}^{\alpha}\rangle$
   lead to the changes of the order of $\langle N_{\alpha}
   \rangle $  in both
   $\langle N^2_{\alpha} \rangle$ and $\langle N_{\alpha}
   \rangle^{2}$, but these changes are the same
   and the correction terms
  are cancelled in the calculation of 
  $\langle \left(\Delta N_{\alpha}\right)^2
  \rangle$. Therefore,
  the macroscopic fluctuations of
  multiplicities are not affected by the CE
  corrections to the average particle numbers. The scaled variances
  of the CE in the thermodynamic limit $V\rightarrow\infty$
  feel the consequences of an exact charge conservation due to the
  suppression of the single mode fluctuations
  $\langle \left(\Delta n_{p}^{\alpha }\right)^2\rangle$ and
  due to the (anticorrelations ) correlations
  between different modes $p\neq k$  with the (same) different
  charge states $\alpha, \beta$. All these effects are absent in the GCE.

%
\section{A system with two conserved
charges}

In the previous sections we have considered the system with one
conserved charge.
 In high energy collisions the measurements of fluctuations
 for the particle numbers and
 (transverse) energies are mainly done for
 electrically charged hadrons. Therefore, in applications
 of the CE
 results to an analysis of the data on fluctuations
 it would be reasonable
 to start with the case when the charge $Q$
 is assumed to be an electric charge.
 On the other hand, other conserved charges are also present in
 the system created in high energy collisions.
In this section we consider the system with two exactly conserved charges
-- electric charge $Q$ and baryonic number $B$. As an example
we study the ideal pion-nucleon gas and neglect 
the quantum statistics effects.
This is the
simplest realistic case where we can study the influence of an
exact $B$ conservation to the
CE fluctuations of the electrically charged particles.
The partition function of this system in the CE is:
\begin{align}\label{ZceQB}
&Z_{c.e.}(V,T,Q,B)
  ~=\!\! \sum_{N_{p},\,N_{\bar{p}}=0}^{\infty}\;
  \sum_{N_{n},\,N_{\bar{n}}=0}^{\infty}\;
     \sum_{N_{\pi},\;N_{\bar{\pi}}=0}^{\infty}
 \frac{\left(\lambda_{p}z_p\right)^{N_{p}}}{N_{p}!}
 \frac{\left(\lambda_{\bar{p}}z_{\bar{p}}
 \right)^{N_{\bar{p}}}}{N_{\bar{p}}!}\;
 \frac{\left(\lambda_{n}z_n\right)^{N_{n}}}{N_{n}!}
 \frac{\left(\lambda_{\bar{n}}
 z_{\bar{n}}\right)^{N_{\bar{n}}}}{N_{\bar{n}}!}\;
 \frac{\left(\lambda_{\pi^+}z_{\pi}\right)^{N_{\pi^+}}}{N_{\pi^+}!}
 \frac{\left(\lambda_{\pi^-}z_{\pi}
 \right)^{N_{\pi^-}}}{N_{\pi^-}!} \nonumber
 \\[0.3cm]
 & \times
  \delta\left[\;\left( N_{p}-
  N_{\bar{p}}\,+\,N_{\pi^+}-N_{\pi^-}\right)-Q\;\right]\;\;
  \delta\left[\;\left( N_{p}-N_{\bar{p}}\,+\,N_{n}-
  N_{\bar{n}}\right)-B\;\right] \nonumber
 \\[0.3cm]
 &\;=\;
 \int_0^{2\pi}\frac{d\varphi}{2\pi}\;\int_0^{2\pi}\frac{d\phi}{2\pi}\;
 \exp\left(\;-i\,Q\,\varphi\, -i\,B\,\phi \;\right)
  \times \exp \left[ z_p\left(\lambda_{p}\,e^{i(\varphi+\phi)}
                   \,+\, \lambda_{\bar{p}}\,e^{-i(\varphi+\phi)}
           \right)\right] \nonumber \\
 &\times\exp\left[z_n \left(\lambda_{n}\,e^{i\phi}
                   \,+\, \lambda_{\bar{n}}\,e^{-i\phi}\right)\right]
            \times\exp\left[ z_{\pi}\,\left(\lambda_{\pi^+}
            \,e^{i\varphi}
                   \,+\, \lambda_{\pi^-}\,e^{-i\varphi}
           \right)
           \right]
             \;=\; \sum_{k=-\infty}^{\infty}
 I_{k-Q}(2z_p)\;I_{k+B-Q}(2z_n)\;I_{k}(2z_{\pi})
        \;,
\end{align}
where we have used that
$\exp[x(t+1/t)]=\sum_{k=0}^{\infty}t^k\;I_k(2x)\;$.
From Eq.\,(\ref{ZceQB}) it follows:
 \eq{
\langle N_{j,\alpha}\rangle_{c.e.}~=~c_{1,\alpha}^j\;z_j\;,~~~
\langle N_{j,\alpha}^2\rangle_{c.e.}~=~c_{1,\alpha}^j\;z_j \;+\;
c_{2,\alpha}^j\;z_j^2\;, \label{NjQB}
}
where $j$ numerates pion, neutron and proton,   $\alpha=1$ corresponds to
particles $\pi^+,~n,~p$ and $\alpha=-1$  to 
antiparticles $\pi^-,~\overline{n},\overline{p}$, and ($m=1,2$) 
\begin{align}\label{cp}
c_{m,\alpha}^p
\;&=\; \sum_{k=-\infty}^{\infty} I_{k+\alpha\cdot m-Q}(2z_p)\;
 I_{k+B-Q}(2z_n)\;I_{k}(2z_{\pi})
 ~\times ~\left[Z_{c.e.}(V,T,Q,B)\right]^{-1} ~,\\
c_{m,\alpha}^n
 \;&=\; \sum_{k=-\infty}^{\infty}
 I_{k-Q}(2z_p)\;I_{k+\alpha\cdot m+B-Q}(2z_n)\;I_{k}(2z_{\pi})
 ~\times ~\left[Z_{c.e.}(V,T,Q,B)\right]^{-1} ~,\label{cn}\\
c_{m,\alpha}^{\pi}
 \;&=\; \sum_{k=-\infty}^{\infty}
 I_{k+\alpha\cdot m-Q}(2z_p)\;I_{k+\alpha\cdot m+B-Q}(2z_n)\;I_{k}(2z_{\pi})
 ~\times ~\left[Z_{c.e.}(V,T,Q,B)\right]^{-1} ~.\label{cpi}
\end{align}
%
%
Formulas
for the cross-averages $\;\langle N_i N_j\rangle\;$ can be obtained in
the similar manner.
The calculations with Eqs.\,(\ref{cp}-\ref{cpi}) are effective
for small systems. In this case the $k$-sums in the above equations
converge rapidly and small number of terms lead
to the accurate results.
%
%
In the limit of large system volume we can use another
technique, similar to that developed in the previous section.
This leads to simple analytical results.
Using this method one can obtain, for example, the
scaled variances for (negatively) positively  charged particles in the
thermodynamic limit.
The same pictures can be obtained directly from Eqs.\,(\ref{cp}-\ref{cpi})
by numerical calculations at
$\;z_p,\;z_n,\;z_{\pi} \gg 1\;$.

\begin{figure}[h!]
\epsfig{file=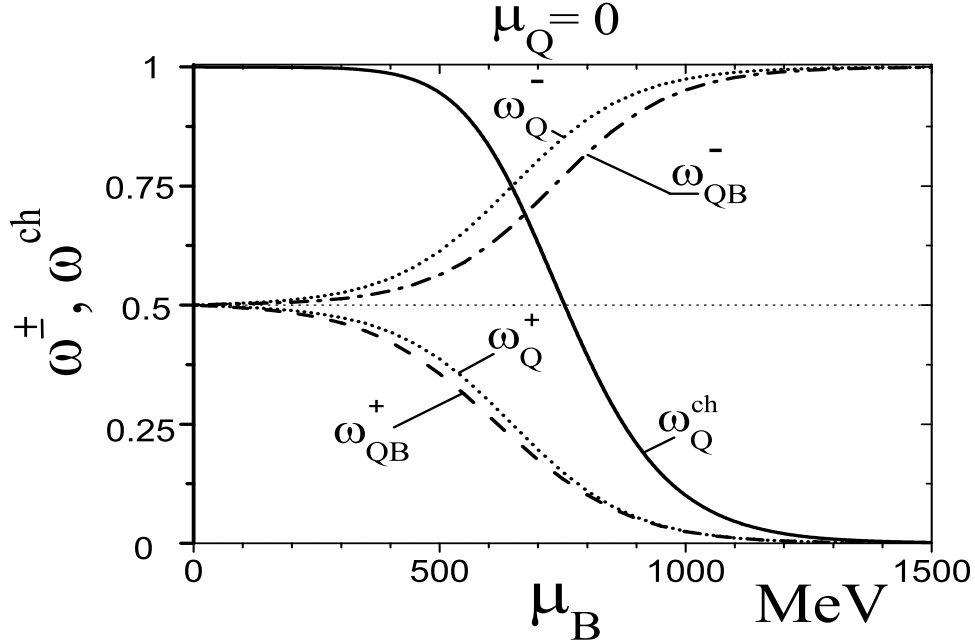,height=9cm,width=13cm}
 \caption{The scaled variances $\omega_{Q,B}^{+}$ (dashed line)
 and $\omega_{Q,B}^{-}$ (dashed-dotted line) given by Eq.\,(\ref{WQBsteph}).
 The dotted lines show the scaled variances
 $\omega_{Q}^{+}$  and $\omega_{Q}^{-}$ given by Eq.\,(\ref{WQsteph}).
 The solid line presents the scaled variance for all charged
 particles $\omega_{Q}^{ch}$.
 The results correspond to $T=120$~MeV.
 }
 \label{WQmceFig}
\end{figure}

First, we consider the case when the electric charge $Q$ is
exactly conserved and the baryonic number $B$
conservation is treated within the GCE. This results in:
\begin{align}\label{WQsteph}
\omega_{Q}^{\pm}
 \;=\; 1\;-\;\frac{z_p^{\pm}+z_{\pi}^{\pm}}{z_px_p+z_{\pi}x_{\pi}}\;,
\end{align}
where
\eq{\;z_j^{\pm}=z_j\exp\left(\pm\frac{\mu_j}{T}\right)\;,~~~
\;x_j=\exp\left(\frac{\mu_j}{T}\right)~+~\exp\left(-\frac{\mu_j}{T}\right)\;,
}
and the chemical potentials $\;\mu_j\;$ are equal to
$\;\mu_p=\mu_Q+\mu_B\;$ for protons
and $\;\mu_{\pi}=\mu_Q\;$ for $\;\pi^+-$mesons.
When both $Q$ and $B$ are exactly conserved,
the CE scaled variances of (negatively) positively  charged particles
are equal to:
\begin{align}\label{WQBsteph}
\omega_{Q,B}^{\pm}
 \;=\; 1\;-\;\frac{z_p^{\pm\,2}(z_nx_n+z_{\pi}x_{\pi})
    + z_{\pi}^{\pm\,2}(z_px_p+z_nx_n) + 2z_p^{\pm}z_{\pi}^{\pm}z_nx_n}
    {(z_p^{\pm}+z_{\pi}^{\pm})~(z_px_pz_nx_n+
    z_px_pz_{\pi}x_{\pi}+z_nx_nz_{\pi}x_{\pi})}\;,
\end{align}
where $x_n=\exp(\mu_B/T)+\exp(-\mu_B/T)$.
Let us repeat that both $\omega_{Q}^{\pm}$ (\ref{WQsteph})
and $\omega_{Q,B}^{\pm}$ (\ref{WQBsteph}) are obtained in the thermodynamic
limit $V\rightarrow\infty$.
The $\omega_{Q}^{\pm}$ (\ref{WQsteph}) corresponds to the CE
for electric charge and the GCE for baryonic number.
The  $\omega_{Q,B}^{\pm}$ (\ref{WQBsteph}) corresponds the CE for both
conserved charges.
It is easy to prove
that $\omega_{Q,B}^{\pm}\leq \omega_{Q}^{\pm}$~, i.e.
an additional exact conservation law reduces
the fluctuations. However, the additional CE suppression of the
scaled variances of (negatively) positively  charged particles due to
the exact
baryonic number conservation is rather small.
We have plotted (\ref{WQsteph}) and (\ref{WQBsteph}) in
Fig.\,\ref{WQmceFig} for $\;\mu_Q=0\;$
 to study the influence of baryon charge conservation
on the fluctuations of electrically charged particles.
%
As one
can see from Fig.\,6 the exact CE baryonic charge conservation
leads to a little additional suppression
and does not change the result $\;\omega^+=\omega^-=0.5\;$ for
zero values of 
the baryonic and electric net charges. Moreover, one can prove that
at zero net charges
any ideal Boltzmann gas with two exactly conserved charges
(i.e. for any
combination of particle charges and their masses) leads
to the scaled variances
equal to
$\omega^{\pm}=0.5$ in the thermodynamic limit.

On the other hand, 
Fig.\,6 demonstrates a strong dependence of the 
$\omega_Q^+$ and $\omega_Q^-$ values 
 on the net baryonic density, it is not
important whether the baryonic number treated within the CE or
the GCE. The matter is that in 
the pion-nucleon gas the electric charge equals to
$\;Q=N_{p}-N_{\bar{p}}\,+ \,N_{\pi^+}-N_{\pi^-}$. At $\mu_B\simeq
0$ the electric charge of the system is close to zero. Then
one finds $\omega_Q^+\simeq\omega_Q^-\simeq 0.5$ (compare to Fig.\,4 at
$y=0.1$). The $\mu_B>0$ leads to $\langle N_{p}\rangle >
\langle N_{\bar{p}}\rangle$, and this means a non-zero electric
charge of the system. In this case an exact electric charge
conservation leads to $\omega_Q^- >\omega_Q^+$ (see Fig.\,4). At
$\mu_B \gg T$ the electric charge density becomes large
due to $\langle N_p\rangle /\langle N_{\bar{p}}\rangle \gg 1$, so
that $\omega_Q^+ \rightarrow 0$ and $\omega_Q^- \rightarrow 1$
(compare to Fig.\,4 at $y=2$).

%
\section{Summary and conclusions}
We have considered the particle number and energy fluctuations for
different systems within the canonical ensemble formulation. The results are
compared to those in the grand canonical ensemble.
We have studied the system with arbitrary number of different
particle species and non-zero net charge in Secs.\,II and III.  An
exact charge conservation reduces the values of  $N_+$ and $N_-$
fluctuations in the thermodynamic limit.
At the non-zero net charge $Q$ the canonical ensemble predicts a
difference for the fluctuations of $N_+$ and $N_-$, they also different
from the fluctuations of all charged particles $N_{ch}=N_++N_-$.
All these features of the canonical ensemble are in a striking difference
with those in the grand canonical ensemble.
 We have demonstrated in Sec.\,IV that the
energy fluctuations of the system are mainly determined by the
fluctuations of the number of particles and have the same volume
dependence. Therefore, the energy fluctuations are rather different
in the canonical and grand canonical ensembles.
We extend our canonical ensemble results and calculate the
 particle number
fluctuations in the system of
single and double charged particles in Sec.\,V,
include the quantum statistics effects in Sec.\,VI, 
study the systems with two conserved
charges in Sec.\,VII.

The canonical ensemble suppression effects for the charged
particle multiplicities are well known, and they are
successfully applied to the statistical description of hadron
production in high energy collisions \cite{ce}. The canonical ensemble
formulation explains, for example, the suppression in a production
of strange hadrons and antibaryons in small systems, i.e. when the
total numbers of strange particles or antibaryons are small. This
consideration demonstrates a difference of the canonical and grand
canonical ensembles -- the statistical ensembles are not
equivalent for small systems. When the size of the system
increases 
all average quantities in both ensembles
become equal. It means that in the 
thermodynamic limit $V\rightarrow\infty$
the canonical ensemble and grand canonical ensemble are equivalent.
Results of Ref.\,\cite{ce-fluc} and the present study demonstrate
that there are also the canonical ensemble effects for the
fluctuations. In contrast to the canonical suppression of average
multiplicities, the canonical effects for the multiplicity
fluctuations do survive at $V\rightarrow\infty$ and they are 
even most clearly
seen in the thermodynamic limit. The changes of the scaled
variances due to an exact charge conservation of the canonical
ensemble are not small (about 50 percent effects) and they are
in general different for positively, negatively and all charged
particles. To observe these new canonical ensemble effects in an
analysis of the data on multiparticle production, several points
should be clarified. To use the condition of an exact charge
conservation one has to apply it to the system of all secondary
hadrons formed in high energy collisions, and this should be done
on the event-by-event basis as we are interested in the system
fluctuations. In the experimental study only a fraction of
produced particles with the conserved charges is detected.
Introducing a probability $q$ that a single particle is accepted
in the detector a
simple relation between the scaled variance of the accepted
particles, $\omega_{acc}$, and the scaled variance of all
particles in the statistical ensemble, $\omega$, was obtained
\cite{ce-fluc}: $\omega_{acc}=q\cdot\omega +(1-q)$. To observe the
real event-byevent fluctuations $\omega$  one needs $q\simeq
1$, otherwise if $q\ll 1$ one always obtains $ \omega_{acc}\simeq
1$ and makes a wrong conclusion that the fluctuations correspond to the
Poisson distribution. 
This fact is of a very general origin, 
and because of relatively small experimental
acceptance a large part
of the event-by-event fluctuations in high energy
multiparticle production is lost.
To observe many interesting event-by-event
fluctuations, for example, due to the QCD critical point (see,
e.g. \cite{step} and references therein), one should accept an
essential part of all secondary particles. In this case the role
of an exact charge conservation would increase. It would also have
a strong influence on an extraction of the so called dynamical
fluctuations (see Ref.\,\cite{fluc1}) from the event-by-event data.

 \begin{acknowledgments}
We would like to thank F.~Becattini, A.I.~Bugrij, T.~Cs\"org\H{o},
M.~Ga\'zdzicki, A.~Ker\"anen, A.P.~Kostyuk, I.N.~Mishustin,
St.~Mr\'owczy\'nski, L.M.~Satarov and Y.M.~Sinyukov for useful
discussions. We thank A.~Swaving for help in the preparation of the
manuscript. The work was supported by US Civilian Research and
Development Foundation (CRDF) Cooperative Grants Program, Project
Agreement UKP1-2613-KV-04.
\end{acknowledgments}

\end{document}